\documentclass{aa} 
\usepackage{graphicx}
\usepackage{multirow}
\newcommand{\bm}[1]{{\mbox{\boldmath $#1$}}}
\usepackage{color}
\usepackage{here}
\usepackage{comment}
\usepackage{lscape}
\usepackage{rotating}

\usepackage{tikz}
\usetikzlibrary{arrows}

\usepackage[colorlinks=true,citecolor=blue]{hyperref}
\usepackage{arydshln}
\usepackage{natbib}

\usepackage{booktabs}% http://ctan.org/pkg/booktabs

\newcommand{\ii}{{{\rm i}}}

\newcommand{\cp}{{c_{\rm p}}}
\newcommand{\cv}{{c_{\rm v}}}
\newcommand{\cs}{C_{\rm s}}

\newcommand{\dd}{{\rm d}}
\newcommand{\betar}{\beta}
\newcommand{\betac}{\beta_{\rm comp}}
\newcommand{\betat}{\beta_{\rm topo}}

\usepackage{ulem}

\newcommand{\tikzmark}[2][-3pt]{\tikz[remember picture, overlay, baseline=-0.5ex]\node[#1](#2){};}

\newcounter{arrow}
\newcommand{\drawcurvedarrow}[3][]{%
 \refstepcounter{arrow}
 \tikz[remember picture, overlay]\draw (#2.center)edge[#1]node[coordinate,pos=0.5, name=arrow-\thearrow]{}(#3.center);
}

\newcommand{\annote}[3][]{%
 \tikz[remember picture, overlay]\node[#1] at (#2) {#3};
}

\usepackage{txfonts}
\begin{document} 

    \title{
%    ---Paper 1---\\
    Theory of solar oscillations in the inertial frequency range:\\
    Linear modes of the convection zone
    }
    
    \titlerunning{Linear model of solar inertial oscillations}

   \author{Yuto Bekki
          \inst{1}
          \and
          Robert H. Cameron\inst{1}
          \and
          Laurent Gizon\inst{1,2,3}
          }

   \institute{Max-Planck-Institut f{\"u}r Sonnensystemforschung,
              Justus-von-Liebig-Weg 3, 37077 G{\"o}ttingen, Germany\\
              \email{bekki@mps.mpg.de}
         \and
         Institut f{\"u}r Astrophysik, Georg-August-Universt{\"a}t G{\"o}ttingen,
         Friedrich-Hund-Platz 1, 37077 G{\"o}ttingen, Germany
         \and
         Center for Space Science, NYUAD Institute,
         New York University Abu Dhabi, Abu Dhabi, UAE
             }

   \date{Received <-->; accepted <-->}

% \abstract{}{}{}{}{} 
% 5 {} token are mandatory
 
  \abstract
  % context 
   {
   On the one hand, several types of global-scale inertial modes of oscillation have been observed on the Sun.
   They include the equatorial Rossby modes, critical-latitude modes, and  high-latitude  modes. 
   On the other hand, the columnar convective modes (predicted by simulations; also known as banana cells or thermal Rossby waves) remain elusive. 
   }
  % aims
   { 
   We aim to investigate the influence of turbulent diffusivities, non-adiabatic stratification, differential rotation, and a latitudinal entropy gradient on the linear global modes of the rotating solar convection zone. 
   }
  % methods 
   {
   We solve numerically for the eigenmodes of a rotating compressible fluid inside a spherical shell.
   The model takes into account the solar stratification, turbulent diffusivities, differential rotation (determined by helioseismology), and the latitudinal entropy gradient.
   As a starting point, we restrict ourselves to a superadiabaticity and turbulent diffusivities that are uniform in space. 
   We identify modes in the inertial frequency range including the columnar convective modes, as well as modes of mixed character.
   The corresponding mode dispersion relations and eigenfunctions are computed for azimuthal orders $m\leq 16$.
     }
  % results 
   { 
   The three main results are as follows.
   Firstly, we find that, for $m \gtrsim 5$, the radial dependence of the equatorial Rossby modes with no radial node ($n=0$) is radically changed from the traditional expectation ($r^m$) for turbulent diffusivities $\gtrsim 10^{12}$~cm$^2$~s$^{-1}$.
   Secondly, we find mixed modes, i.e. modes that share properties of the equatorial Rossby modes with one radial node ($n=1$) and the columnar convective modes, which are not substantially affected by turbulent diffusion.
   Thirdly, we show that the $m=1$ high-latitude mode in the model is consistent with the solar observations when the latitudinal entropy gradient corresponding to a thermal wind balance is included (baroclinally unstable mode).
   }
  % conclusions 
   {
   To our knowledge, this work is the first realistic eigenvalue calculation of the global modes of the rotating solar convection zone. 
   This calculation reveals a rich spectrum of modes in the inertial frequency range, which can be directly compared to the observations. 
   In turn, the observed modes can inform us about the solar convection zone.
   }


   \maketitle
%
%________________________________________________________________
%__________________________________________________________________
%________________________________________________________________
%__________________________________________________________________

%________________________________________________________________
%__________________________________________________________________
%________________________________________________________________

%\tableofcontents

\section{Introduction}

Using 10 years of observations from the Helioseismic and Magnetic Imager (HMI) onboard the Solar Dynamics Observatory (SDO), \citet{gizon2021} discovered that the Sun supports a large number of global modes of inertial oscillations. 
The restoring force for these inertial modes is the Coriolis force, and thus the modes have periods comparable to the solar rotation period ($\sim 27$ days).
The inertial modes can potentially be used as a tool to probe the interior of the Sun, because they are sensitive to properties of the deep convection zone that the p modes are insensitive to. 
In order to achieve this goal, we need a better understanding of the mode physics.

\subsection{Solar inertial modes}

The low frequency modes of solar oscillation have been described in a rotating frame (angular velocity $\Omega_{\rm ref}$). 
Because the Sun is essentially symmetric about its rotation axis, the velocity of each mode in the rotating frame has the form ${\bm{v}} (r,\theta) \exp{[\ii(m\phi-\omega t)]}$, where $r$ is the radius, $\theta$ is the colatitude, $\phi$ is the longitude, $m$ is the azimuthal order, and $\omega$ is the mode eigenfrequency. 
\citet{gizon2021} provide all observed eigenfrequencies $\omega$ for each $m$, and the eigenfunctions ($v_\theta$ and $v_\phi$ at the surface) for a few selected modes.

The first family of inertial modes observed on the Sun consists of the quasi-toroidal  equatorial Rossby modes \citep{loeptien2018}. 
They are analogous to the sectoral r modes described  by, e.g., \citet{papaliozou1978}, \citet{smeyers1981}, and \citet{saio1982}. 
On the Sun these modes have  $3\leq m \leq15$ with a well-defined dispersion relation close to $\omega=-2\Omega_{\mathrm{ref}}/(m+1)$,  where $\omega$ is the mode angular frequency and $\Omega_{\mathrm{ref}}/2\pi=453.1$ nHz is the equatorial rotation rate at the surface. 
For positive $m$, a negative $\omega$ indicates retrograde propagation.
There have been several follow-up studies that confirm these observations \citep[e.g.,][]{liang2019,hanasoge2019,proxauf2020,hanson2020}.
%\citep[e.g.,][]{liang2019,hanasoge2019,proxauf2020, mandal2020,hanson2020,mandal2021,gizon2021}.
Using a one-dimensional $\beta$-plane model with a parabolic shear flow and  viscosity, \citet{gizon2020} show that these modes, among others, are affected by differential rotation and are trapped between the critical latitudes where the phase speed of a mode is equal to the local rotational velocity. 
\citet[][]{fournier2022} extended this model to a spherical geometry using a realistic differential rotation model and found that some Rossby modes can be unstable for $m\leq 3$.

\citet{gizon2021} also report a family of modes at  mid-latitudes that are localized near their critical latitudes. 
Several tens of critical-latitude modes have been identified in the range $m \leq 10$.
Another family of inertial modes introduced by the Sun's differential rotation are the high-latitude modes \citep{gizon2021}. 
The highest amplitude mode ($\sim$ 10~-~20~m~s$^{-1}$ above $50^\circ$ latitude) is the $m=1$ mode with north-south antisymmetric longitudinal velocity $v_\phi$ with respect to the equator. 
This $m=1$ mode was identified by \citet{gizon2021} using linear calculations in two-dimensional model, which are further discussed in the present paper and \citet{fournier2022}.
It corresponds to the spiral-like velocity feature reported at high latitudes by \citet{hathaway2013}, although it was there reported as giant-cell convection.

The equatorial-Rossby and high-latitude modes involve mostly toroidal motions with a radial velocity which is small compared to the horizontal velocity components.
Non-toroidal inertial modes have also been theoretically studied, mainly for incompressible fluids.
These modes tend to be localized onto so-called attractors, closed periodic orbits of rays reflecting off the spherical boundaries \citep[][]{maas1995,rieutord1997,rieutord2001,rieutord2018,sibgatullin2019}. 
They are also strongly affected by critical latitudes when differential rotation is included \citep[e.g.,][]{baruteau2013,guenel2016}.

\subsection{Columnar convective modes}

In numerical simulations of solar-like rotating convection, equatorial convective columns aligned with the rotation axis are prominent \citep[e.g.,][]{miesch2008,bessolaz2011,matilsky2020}. 
They are known as ``Busse columns'' \citep[after][]{busse1970}, or ``thermal Rossby waves'', or ``banana cells'' in the literature.
We call them  ``columnar convective modes'' in the rest of this paper.
These convective columns propagate in the prograde direction owing either to the ``topographic $\beta$-effect'' originating from the geometrical curvature  \citep[e.g.,][]{busse2002} or to the ``compressional $\beta$-effect'' originating from the strong density stratification \citep[][]{ingersoll1982,evonuk2008,glatzmaier2009,evonuk2012,verhoeven2014}.
\citet{glatzmaier1981} numerically derived the dispersion relation and the radial eigenfunctions of these convective modes using a one-dimensional cylinder model.
They showed that the fundamental ($n=0$) mode is the fastest of these prograde propagating modes with an eigenfunction that is localized near the surface, where the compressional $\beta$-effect is strongest.

In the parameter regime of the various numerical simulations, the columnar convective modes are the structures that are the most efficient to transport thermal energy upward under the rotational constraint \citep[e.g.,][]{gilman1986,miesch2000,brun2004,miesch2008,kapyla2011,gastine2013,hotta2015,featherstone2016,matilsky2020,hindman2020}.
Furthermore, it is often argued that these convective modes play a critical role in transporting the angular momentum equatorward to maintain the differential rotation of the Sun \citep[e.g.,][]{gilman1986,miesch2000,balbus2009}.
The dominant columnar convective modes seen in simulations have not been detected in the velocity field at the surface of the Sun. 
However, we will show in this paper that some retrograde inertial modes have a mixed character and share some properties with columnar convection.

%__________________________________________________________________
%__________________________________________________________________
\subsection{Focus of this paper}

In this paper, we study the properties of the equatorial Rossby modes, the  high-latitude inertial modes, and the columnar convective modes in the linear regime.
We are mainly interested in the effects of turbulent diffusion, solar differential rotation, and non-adiabatic stratification on these modes.
Note that the critical-latitude modes, which are discussed by \citet[][]{fournier2022}, will not be dealt with in depth in this paper.

Firstly, we will show that, when the turbulent viscosity is above approximately $10^{12}$~cm$^2$ s$^{-1}$, the equatorial Rossby modes with no radial node ($n=0$) strongly depart from the expected $r^m$ dependence and the radial vorticity at the surface is no longer maximum at the equator at azimuthal wavenumbers $m \gtrsim 5$.
Secondly, we report a new class of modes with frequencies close to that of the classical Rossby modes.
They share properties of both equatorial Rossby modes and convective modes.
Thirdly, we provide a physical explanation for the properties of the $m=1$ high latitude modes in terms of the baroclinic instability due to the latitudinal entropy gradient in the convection zone.

The organization of the paper is as follows.
In \S \ref{sec:model} we specify the linearized equations and solve the eigenvalue problem. 
The low-frequency modes are discussed in \S\ref{sec:results} for the inviscid, adiabatically stratified, and uniformly-rotating case. 
Then, the effects of turbulent diffusion and a non-adiabatically stratified background are discussed in \S\ref{sec:viscosity} and  \S\ref{sec:delta}.
We discuss how the solar differential rotation and the associated baroclinicity affect the mode properties in \S\ref{sec:diffrot}.
The results are summarized  in \S\ref{sec:summary}.

%__________________________________________________________________
%__________________________________________________________________

\section{Eigenvalue problem} \label{sec:model}

In order to investigate the properties of various inertial modes in the Sun, a new numerical code has been developed.
We consider the linearized fully-compressible hydrodynamic equations in a spherical coordinate $(r,\theta,\phi)$.

%__________________________________________________________________

\subsection{Linearized equations} \label{sec:eqs}

The linearized equations of motion, continuity, and energy conservation are:
\begin{eqnarray}
&& \frac{\partial \bm{v}}{\partial t}=-\frac{\nabla p_{1}}{\rho_{0}}-\frac{\rho_{1}}{\rho_{0}}g\bm{e}_{r}- (\Omega-\Omega_{0})\frac{\partial\bm{v}}{\partial\phi} 
	-2\Omega\bm{e}_{z}\times\bm{v} \nonumber \\ 
&& \ \ \ \ \ \ \ \ \  -r\sin{\theta}\ \bm{v}\cdot\nabla \Omega+\frac{1}{\rho_{0}}\nabla\cdot \bm{\mathcal{D}},  \label{eq:vel} \\
&& \frac{\partial \rho_{1}}{\partial t}=-\nabla\cdot(\rho_{0}\bm{v})- (\Omega-\Omega_{0})\frac{\partial \rho_{1}}{\partial\phi},  \label{eq:mass} \\
&&\frac{\partial s_{1}}{\partial t}=\cp \delta\frac{v_{r}}{H_{p}}
    -\frac{v_{\theta}}{r}\frac{\partial s_{0}}{\partial \theta}
	- (\Omega-\Omega_{0})\frac{\partial s_{1}}{\partial \phi} \nonumber \\
&& \ \ \ \ \ \ \ \ \  +\frac{1}{\rho_{0}T_{0}}\nabla\cdot(\kappa\rho_{0}T_{0}\nabla s_{1}) , \label{eq:ent}
\end{eqnarray}
where, $\bm{v}=(v_{r},v_{\theta},v_{\phi})$ is the 1st-order velocity perturbation. 
In this paper, we only consider the differential rotation for the mean flow and ignore meridional circulation.
Thus, the background velocity is $\bm{U}=r\sin{\theta} (\Omega-\Omega_{0})\bm{e}_{\phi}$.
Here, $\Omega$ is a function of $r$ and $\theta$ and denotes the rotation rate in the Sun's convection zone, and $\Omega_{0}$ is the rotation rate of the observer's frame.
Note that, in this paper, we start our by analysing the case without differential rotation for simplicity and study the linear modes in the uniformly-rotating Sun.
In this case, $\Omega_{0}$ represents the rotation rate of the unperturbed background state.
For the case with the solar differential rotation, we choose to use the Carrington rotation rate $\Omega_{0}/2\pi=456.0$ nHz.

The unperturbed model is given by $p_{0}$, $\rho_{0}$, $T_{0}$, $g$, and $H_{p}$  which are the pressure, density, temperature, gravitational acceleration, and pressure scale height of the background state.
The background is assumed to be spherically symmetric and in an adiabatically-stratified hydrostatic balance. 
All of these variables are functions of $r$ alone. 
We use the same analytical model as \citet{rempel2005} and \citet{bekki2017a} for the background stratification which nicely mimics the solar model~S \citep{jcd1996}.
The variables with subscript $1$, $p_{1}$, $\rho_{1}$, and $s_{1}$, represent the 1st-order perturbations of pressure, density, and entropy that are associated with velocity perturbation $\bm{v}$.
Here, to close the equations, the linearized equation of state is used
\begin{eqnarray}
&&  \frac{p_{1}}{p_{0}}=\gamma\frac{\rho_{1}}{\rho_{0}}+\frac{s_{1}}{\cv},
\end{eqnarray}
where $\gamma=5/3$ is the specific heat ratio and $\cv$ denotes the specific heat at constant volume. 

Although the background is approximated to be adiabatic, we can still introduce a small deviation from the adiabatic stratification in terms of the superadiabaticity $\delta=\nabla-\nabla_{\mathrm{ad}}$, where $\nabla=\dd \ln T/ \dd \ln p$ is the double-logarithmic temperature gradient.
In the solar convection zone, superadiabaticity is estimated as $\delta\approx10^{-6}$ \citep[e.g.,][]{ossendrijver2003}. 
Also, when the solar differential rotation is included, we may add a latitudinal entropy variation $\partial s_{0}/\partial \theta$ that is associated with the thermal wind balance of the differential rotation \citep[e.g.,][]{rempel2005,miesch2006,brun2011}.

We assume that the viscous stress tensor, $\bm{\mathcal{D}}$, is given by
\begin{eqnarray}
&& \mathcal{D}_{ij}=\rho_{0}\nu\left( \mathcal{S}_{ij}-\frac{2}{3}\delta_{ij}\ \nabla\cdot\bm{v}\right),
\end{eqnarray}
where $\mathcal{S}$ is the deformation tensor. 
See \citet{fan2014} (their equations 8 to 13) for detail expressions of $\mathcal{S}_{ij}$ in  spherical coordinates.
The viscous and thermal diffusivities are denoted by $\nu$ and $\kappa$ respectively.

%%%%%%%%%%%%%%%%%%%%%%%%%%%%%%%%%%%%%%%%%%%%%%%
\subsection{Eigenvalue problem} \label{sec:eigenvalue}

%___________________________________________________________
   \begin{figure}
     \centering
     \includegraphics[width=0.55\linewidth]{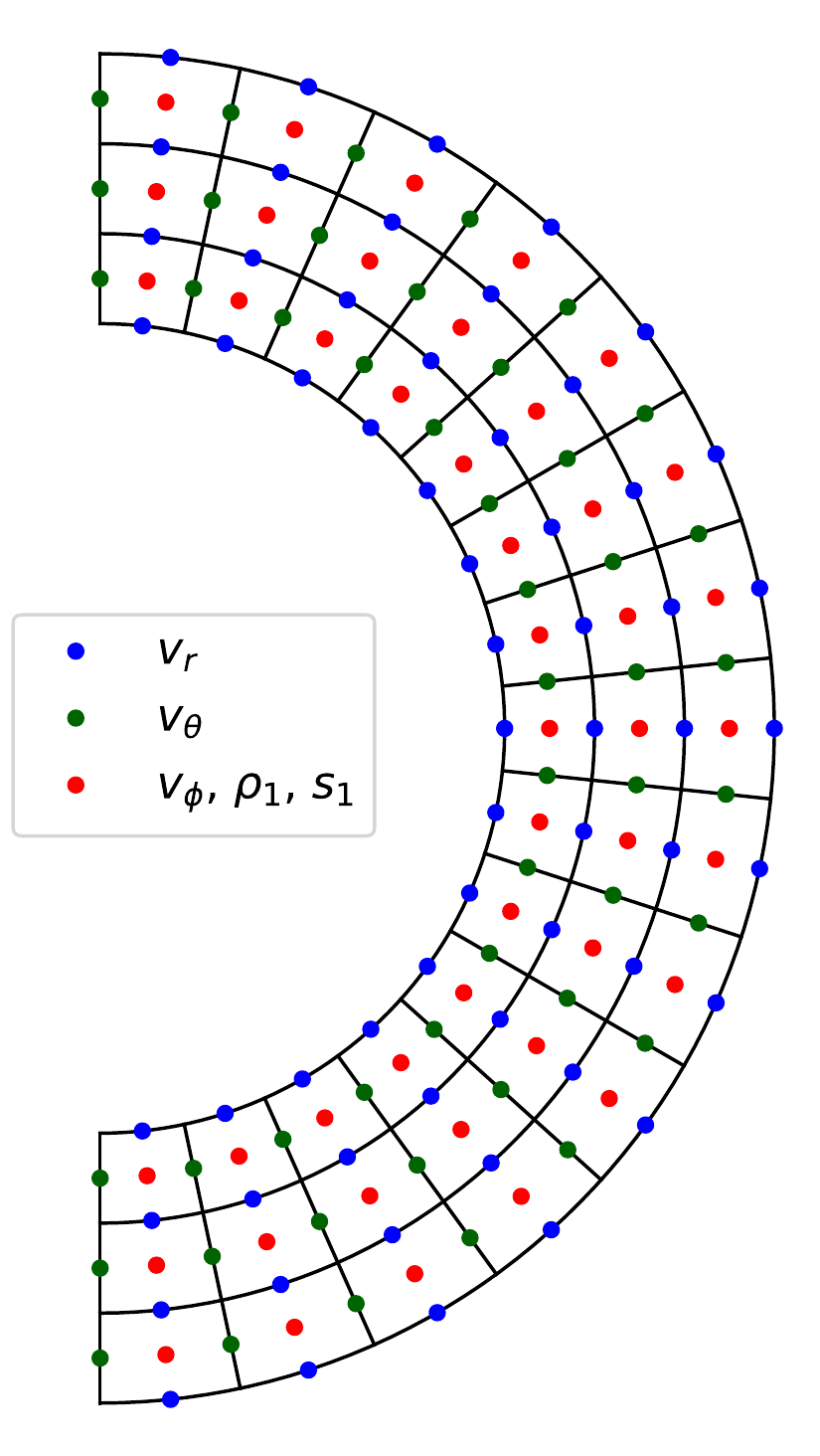}
     \caption{
     Layout of the staggered grid used to solve the eigenvalue equation. 
     The grid locations where $v_{\phi}$, $\rho_{1}$, and $s_{1}$ are defined are denoted by red circles.
     The blue and green circles represent the grid locations of $v_{r}$ and $v_{\theta}$, respectively.
     The grid resolution is reduced for a visualization purpose.}
     \label{fig:grid}
   \end{figure}
%___________________________________________________________

We assume that the $\phi$ and $t$ dependence of all the perturbations $\bm{v}$, $\rho_{1}$, $p_{1}$, and $s_{1}$ is given by the waveform $\exp{[\ii(m\phi-\omega t)]}$, where $m$ is the azimuthal order (an integer) and $\omega$ is the complex angular frequency.
With this representation, Eqs.~(\ref{eq:vel})--(\ref{eq:ent}) give
\begin{eqnarray}
&&\omega v_{r} = -\ii \frac{\partial}{\partial r}\left[\cs^2 \left(  \frac{\rho_1}{\rho_0}+ \frac{s_{1}}{\cp}\right)\right]+\ii \frac{g }{\cp} s_{1}
	+2\ii \Omega\sin{\theta} v_{\phi} 
	\nonumber \\
&& \ \ \ \ \ \ \ \ \  	+m(\Omega-\Omega_{0})v_{r} +\frac{\ii}{\rho_{0}}(\nabla\cdot \mathcal{D})_{r},  
\label{eq2:vr} 
\\
&& \omega v_{\theta}=-\frac{\ii}{r}\frac{\partial}{\partial \theta}\left[\cs^{2}\left({ \frac{\rho_1}{\rho_0}}+\frac{s_{1}}{\cp}\right)\right]+2\ii \Omega\cos{\theta}\ v_{\phi} 
\nonumber \\
&& \ \ \ \ \ \ \ \ \  	+m(\Omega-\Omega_{0})v_{\theta} +\frac{\ii}{\rho_{0}}(\nabla\cdot \mathcal{D})_{\theta},  
\label{eq2:vq} 
\\
&& \omega v_{\phi} = -\frac{m \cs^{2}}{r\sin{\theta}}\left({ \frac{\rho_1}{\rho_0}}+ \frac{s_{1}}{\cp}\right)- 2\ii \Omega(v_{r}\sin{\theta}+v_{\theta}\cos{\theta}) 
\nonumber \\
&& \ \ \ \ \ \ \ \ \ 	+m(\Omega-\Omega_{0}) v_{\phi}- \ii r\sin{\theta}\left(v_{r}\frac{\partial \Omega}{\partial r}
	+ \frac{v_{\theta}}{r}\frac{\partial \Omega}{\partial\theta} \right) 
	\nonumber \\
&& \ \ \ \ \ \ \ \ \  	+\frac{\ii}{\rho_{0}}(\nabla\cdot {\mathcal{D}})_{\phi},  \label{eq2:vp} \\
&& \omega  {\rho_1}=-\ii{\rho_0}\nabla\cdot\bm{v}+\ii \frac{\rho_0}{H_{\rho}} v_{r}+ m(\Omega-\Omega_{0}) \rho_1,  
 \label{eq2:rho} 
 \\
&& \omega  s_{1}= \ii \frac{ \cp \delta }{H_{p}}v_{r} -\frac{\ii}{r}\frac{\partial s_{0}}{\partial \theta} v_{\theta}+ m(\Omega-\Omega_{0})s_{1} 
\nonumber \\
&& \ \ \ \ \ \ \ \ \   	-\frac{\ii}  {\rho_{0}T_{0}}\nabla\cdot\left(\kappa\rho_{0}T_{0}\nabla s_1\right),
\label{eq2:ent}
\end{eqnarray}
where $\cs=(\gamma p_{0}/\rho_{0})^{1/2}$ is the sound speed and $\cp=\gamma \cv$ is the constant specific heat at constant pressure.
Here, the longitudinal velocity $v_{\phi}$, density perturbation $\rho_{1}$, and entropy perturbation $s_{1}$ are out of phase with the meridional components of  velocity ($v_{r}$ and $v_{\theta}$) in the inviscid limit ($\nu=\kappa=0$).

Equations (\ref{eq2:vr})--(\ref{eq2:ent}) can be combined into an eigenvalue problem
\begin{eqnarray}
&&\omega \bm{V}=M\bm{V}, \label{eq:eigen}
\end{eqnarray}
where 
\begin{eqnarray}
&& \bm{V}=\left(
     	\begin{array}{c}
		v_{r} \\
		v_{\theta} \\
		v_{\phi} \\
		\rho_1 \\
		s_1
	\end{array}
    \right)
\end{eqnarray}
and $M$ is the linear differential operator represented by the right-hand side of the Eqs.~(\ref{eq2:vr})--(\ref{eq2:ent}).
The operator $M$ depends on azimuthal order $m$ and the model parameters such as differential rotation $\Omega(r,\theta)$, superadiabaticity $\delta$, and diffusivities $\nu$ and $\kappa$.

%___________________________________________________________
   \begin{figure*}[h]
     \centering
     \includegraphics[width=\linewidth]{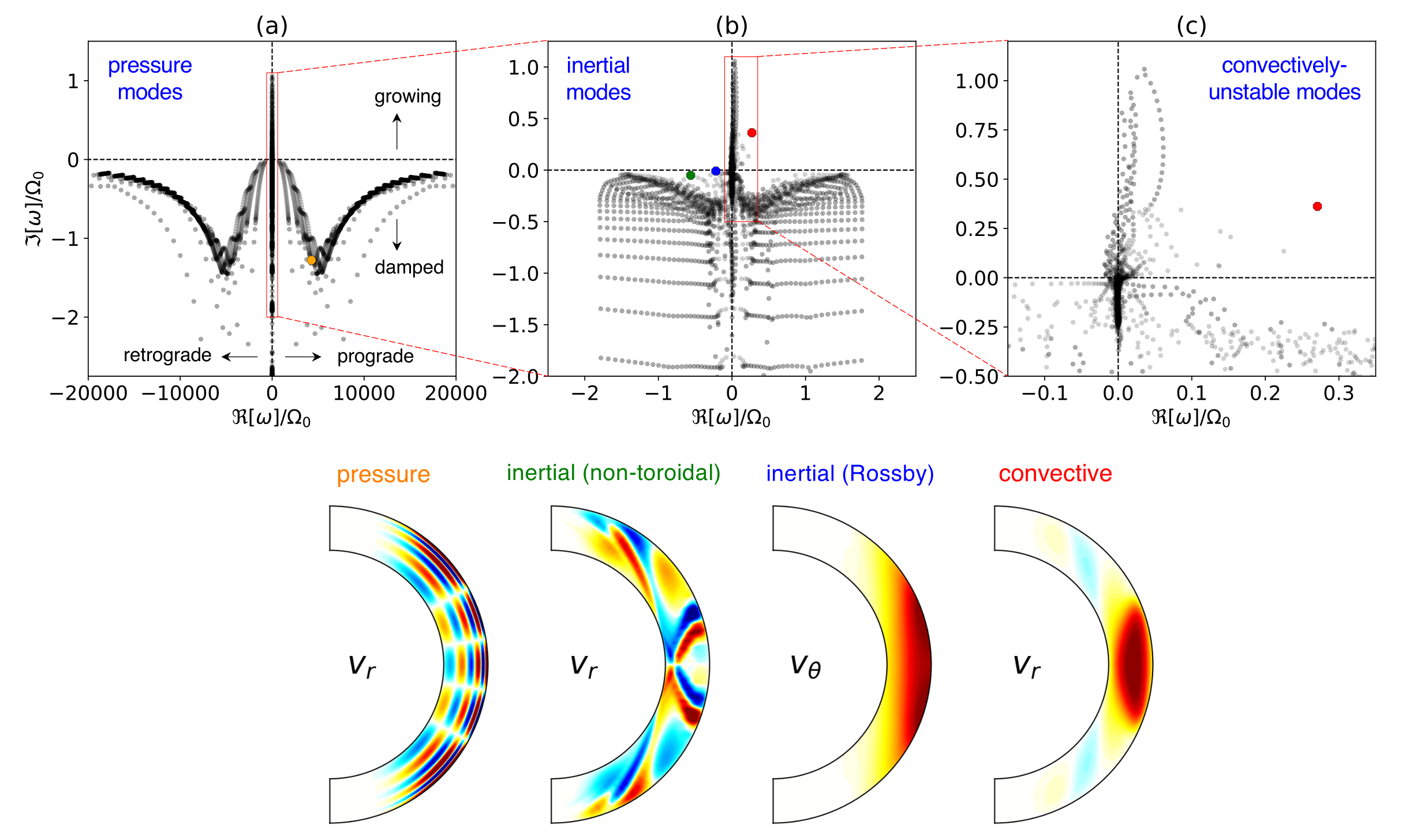}
     \caption{
     Upper panels: Complex eigenfrequencies $\omega$ in the co-rotating frame for $m=8$ in the case of uniform rotation 
     ($\Omega=\Omega_0$), a weakly superadiabatic stratification ($\delta=10^{-6}$), and moderate turbulent viscous and thermal diffusivities ($\nu=\kappa=10^{11}$ cm$^{2}$ s$^{-1}$).
     Panel (a): Real frequencies in the range $\pm 10$~mHz showing the acoustic modes (p modes).
     Panel (b): Zoom-in focusing on the inertial range $|\Re[\omega]|<2\Omega_{0}$.
     Panel (c): Zoom-in focusing on the convectively-unstable modes ($\Im[\omega]>0$).
     Lower panels: Example eigenfunctions of pressure (acoustic), non-toroidal inertial, toroidal inertial (equatorial Rossby), and columnar convective modes, from left to right.
     The eigenfrequencies of these modes are highlighted by orange, green, blue, and red dots in the upper panels.
     }
     \label{fig:allfreq}
   \end{figure*}
%__________________________________________%
%------------------------------------------%
\begin{figure*}[]
\centering
    \includegraphics[width=0.75\linewidth]{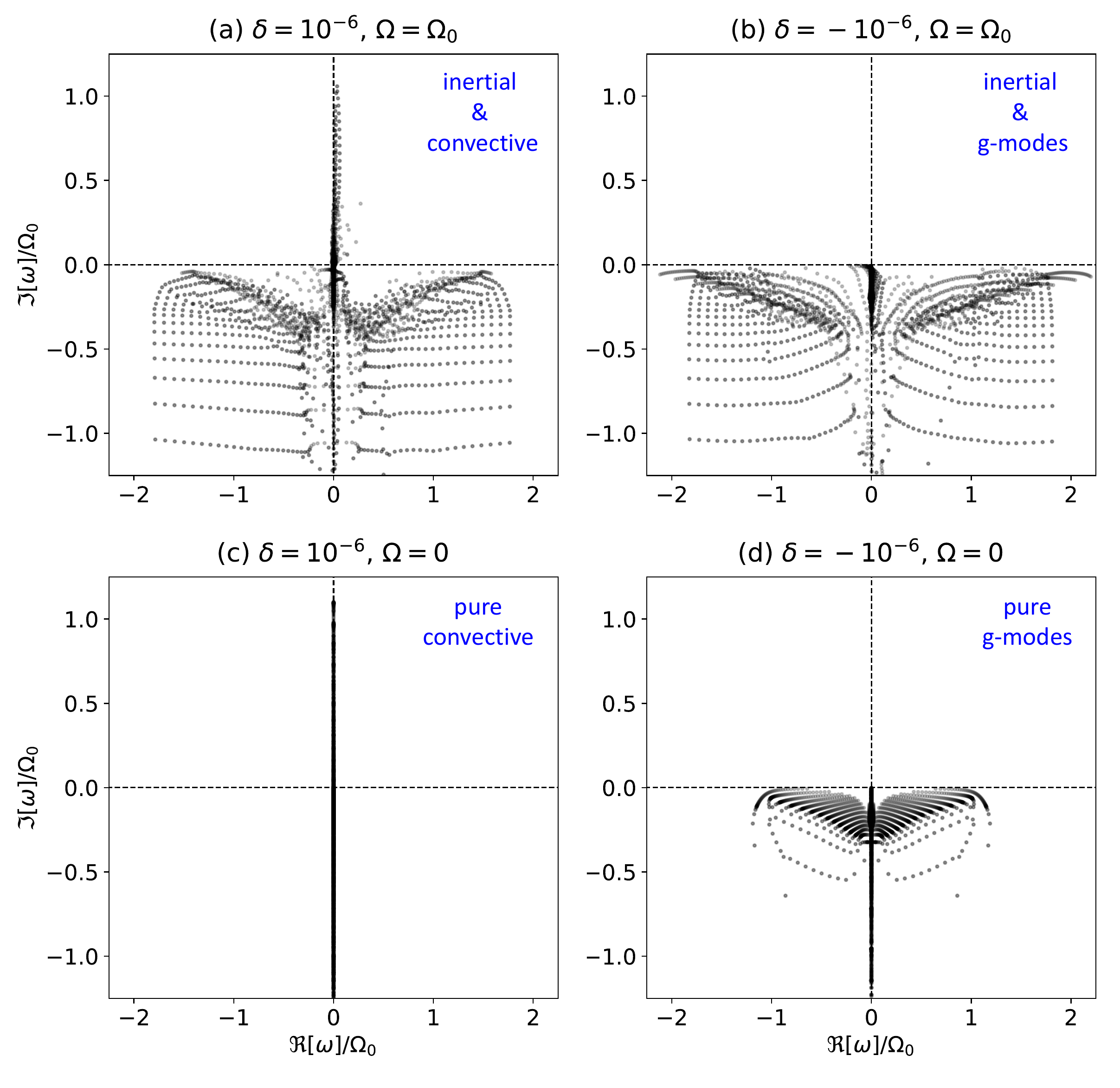}
    \caption{
    Eigenfrequency spectrum in the complex plane at $m=8$ for (a) $\delta=10^{-6}$, $\Omega=\Omega_0$, (b) $\delta=-10^{-6}$, $\Omega=\Omega_0$, (c) $\delta=10^{-6}$, $\Omega=0$, and (d) $\delta=-10^{-6}$, $\Omega=0$, respectively.
    Here, $\Omega_0$  is the Carrington rotation rate.
    Only inertial frequency range is shown.
    Upper and lower panels show the cases with and without uniform rotation.
    Left and right panels show the cases with superadiabatic and subadiabatic background.
    Panel (a) is the same as Fig.~\ref{fig:allfreq}b.
    }
    \label{fig:freqzoom_supomega}
\end{figure*}
%------------------------------------------%

%%%%%%%%%%%%%%%%%%%%%%%%%%%%%%%%%%%%%%%%%%%%%%%
\subsection{Boundary conditions} \label{sec:bc}

In this study, we confine our numerical domain from $r_{\mathrm{min}}=0.71R_{\odot}$ to $r_{\mathrm{max}}=0.985R_{\odot}$ in the radial direction to avoid the strong density stratification near the solar surface and gravity modes in the radiative interior. 
Because of viscosity, in this problem we have four second-order (in both the radial and latitudinal directions) PDEs and one first-order PDE. 
Equation~(\ref{eq2:rho}) does not increase the order of the system as $\rho_1$ can be eliminated from the system  without increasing the order of the other equations.
Thus eight boundary conditions are required in the radial direction (four at the top, four at the bottom).
At the top and bottom, we use impenetrable horizontal stress-free conditions for the velocity and assume there is no entropy flux ($\propto  \kappa \partial s_{1}/\partial r$) across the boundary:
\begin{eqnarray}
&& v_{r}=0, \ \ \ \frac{\partial}{\partial r}\left( \frac{v_{\theta}}{r}\right)=\frac{\partial}{\partial r}\left( \frac{v_{\phi}}{r}\right)=0, \ \ 
\frac{\partial s_1}{\partial r}=0.
\end{eqnarray}
All latitudes are covered in the numerical scheme, from the north pole ($\theta=0$) to the south pole ($\theta=\pi$). 
We need another eight boundary conditions in the $\theta$ direction.
For non-axisymmetric cases ($m\neq 0$), at the poles we impose 
\begin{eqnarray}
&& v_{r}=v_{\theta}=v_{\phi}=0, \ \  s_1=0,
\end{eqnarray}
to make the quantities single valued.
For the axisymmetric case ($m=0$), at both poles we assume instead
\begin{eqnarray}
&& \frac{\partial v_{r}}{\partial \theta}=v_{\theta}=0, \ \ \ 
\frac{\partial}{\partial\theta}\left(\frac{v_{\phi}}{\sin{\theta}} \right)=0, \ \ \frac{\partial s_1}{\partial \theta}=0.
\end{eqnarray}

%-------------------------------------------------------%
\subsection{Numerical scheme} \label{sec:numerical}

We numerically solve the above eigenvalue problem  using a finite differencing method in the meridional plane. 
We use a spatially-uniform grids.
The grids for $v_{\phi}$, $\rho_1$, and $s_1$ are staggered grids by half a grid point in radius for $v_{r}$ and half a grid point in colatitude for $v_{\theta}$ \citep[following][]{gilman1975}, as is illustrated in Fig.~\ref{fig:grid}.
Spatial derivatives are evaluated with a centered second-order accurate scheme.
By converting the two dimensional grid ($N_{r}, N_{\theta}$) into one dimensional array with the size $N_{r}N_{\theta}$ for all variables, $\bm{V}$ is defined as a one dimensional vector with size $\sim 5N_{r}N_{\theta}$.
Once the boundary conditions are properly set, ${M}$ can be constructed as a two-dimensional complex matrix with the size  approximately ($5N_{r}N_{\theta}\times5N_{r}N_{\theta}$).
This method is similar to that of \citet{guenther1985}.
In practice, each element of ${M}$ can be computed by substituting a corresponding unit vector $\bm{V}$ into the right-hand side of the Eqs.~(\ref{eq2:vr})--(\ref{eq2:ent}). 
In most of the calculations, we use the grid resolution of $(N_{r},N_{\theta})=(16,72)$.
We have also carried out higher-resolution calculations with $(N_{r},N_{\theta})=(24,180)$ for a uniform rotation case to check the grid convergence of the results. 
When the grid resolution is increased, the total number of eigenmodes increases accordingly. 
The additional modes have higher radial and latitudinal wavenumbers and are more finely structured. 
For the interpretation of the large-scale modes which have been observed on the Sun, the results are converged with $(N_{r},N_{\theta})=(16,72)$.

We use the LAPACK routines \citep[][]{lapack99} to numerically compute the eigenvalues and eigenvectors of ${M}(m,\nu,\kappa,\delta,\Omega)$,
corresponding to the mode frequencies $\omega$ and the eigenfunctions $(v_{r},v_{\theta},v_{\phi},\rho_{1},s_{1})$ of linear modes in the Sun.
In this study, we limit the range of azimuthal orders to $m\geq 0$ and allow the real frequency to take a negative value.
This means that $\Re{[\omega]}<0$ corresponds to retrograde-propagating modes and $\Im{[\omega]}>0$ corresponds to exponentially growing modes.

 %%%%%%%%%%%%%%%%%%%%%start table%%%%%%%%%%%%%%%%%%%%%%%%%%%%
\begin{table*}[h]
  \begin{center} 
\caption{Summary of the properties of the modes of the models discussed in this paper. Each row refers to a set of modes with different $m$ values.} \label{table:summary}
\small
\begin{tabular}{cccccccccccc} 
\toprule
\toprule
 \multirow{3}{*}{Classification}  & & {peak location} & \multicolumn{3}{c}{north-south} &  propagation & sections   \\
  & & of kinetic energy & \multicolumn{3}{c}{symmetries}  & direction & discussed \\
  & &  & $v_r$&$v_\theta$&$v_\phi$ &\\ 
 \midrule
 \midrule
 Equatorial Rossby ($n=0$) & &  equator  & A & S & A & retrograde & \S~\ref{sec:n=0}, \S~\ref{sec:viscosity}, \S~\ref{sec:viscouscrit}  \\
\tikzmark[xshift=-1.75em]{a}   Equatorial Rossby ($n=1$) & &  equator  & A&S&A  & retrograde & \S~\ref{sec:n=1}, \S~\ref{sec:viscouscrit}  \\
 Columnar convective ($\zeta_{z}$-sym) & &  equator  & S& A& S & prograde & \S~\ref{sec:thNS-S}, \S~\ref{sec:delta}  \\
 
 \tikzmark[xshift=-0.5em]{b}  Columnar convective ($\zeta_{z}$-antisym) & &  equator  & A &S& A  & prograde & \S~\ref{sec:thNS-AS}  \\
 High latitude ($\zeta_{z}$-sym) & &  near poles  & S&A&S  & retrograde & \S~\ref{sec:tpgNS-S}  \\
 High latitude ($\zeta_{z}$-antisym) & &  near poles  &A&S&A  & retrograde & \S~\ref{sec:tpgNS-AS}, \S~\ref{sec:highlat_m1}  \\
\bottomrule
\end{tabular}
\drawcurvedarrow[bend left=-60,stealth-stealth]{a}{b}
\annote[left]{arrow-1}{``mixed''}
\end{center}
\vspace{-1.0\baselineskip}
\tablefoot{
The integer $n$ denotes the number of radial nodes of $v_{\theta}$ at the equator for the Rossby modes.
The north-south symmetries of the different components of the velocity 
are given in columns 3 to 6, where `S' indicates the velocity component is symmetric across the equator and `A' indicates the velocity component is antisymmetric across the equator.
The propagation direction is for the uniformly rotating case, and is given in the rotating frame.
%The propagation frequency $\Re[\omega]$ shown in 5th column represents the mode frequency for the case of uniform rotation.
%\LG{what does the last sentence mean?}
} 
 \end{table*}
%%%%%%%%%%%%%%%%%%%%%%%%end table%%%%%%%%%%%%%%%%%%%%%%%%%%%%

%%%%%%%%%%%%%%%%%%%%%start table%%%%%%%%%%%%%%%%%%%%%%%%%%%%
\begin{table*}[h] 
 \begin{center} 
\caption{Dispersion relations of the modes of the model with uniform rotation ($\Omega=\Omega_{0}$), 
$\nu=\kappa=0$, and $\delta=0$. Frequencies are measured in the corotating frame.} 
\label{table:disp}
\small
\begin{tabular}{ccccccccccccc} 
\toprule
\toprule
  \renewcommand{\arraystretch}{1.8}
  \multirow{2}{*}{$m$} & & \multicolumn{8}{c}{$\Re[\omega]/\Omega_{0}$} 
  \\ \cmidrule{3-10}
\multicolumn{2}{c}{ } &\multicolumn{2}{c}{Equatorial Rossby modes}& &  \multicolumn{2}{c}{Columnar convective modes} &  & \multicolumn{2}{c}{High-latitude modes} \\
 	 & & $n=0$ & $n=1$ & & $\zeta_{z}$ sym. & $\zeta_{z}$ antisym. & & $\zeta_{z}$ sym. & $\zeta_{z}$ antisym. \\
\cmidrule{3-4} \cmidrule{6-7}  \cmidrule{9-10}
$0$  & &  --       & $-0.629$ & & --      &$0.629$ & & --    & -- \\
$1$  & &$-0.999$  & $-0.527$ & &$0.151$ &$0.694$ & &$-0.303$ &$-0.173$ \\
$2$  & &$-0.666$  & $-0.447$ & &$0.290$ &$0.758$ & &$-0.293$ &$-0.172$ \\
$3$  & &$-0.499$  & $-0.380$ & &$0.410$ &$0.824$ & &$-0.258$ &$-0.166$ \\
$4$  & &$-0.399$  & $-0.328$ & &$0.518$ &$0.883$ & &$-0.216$ &$-0.157$ \\
$5$  & &$-0.333$  & $-0.286$ & &$0.612$ &$0.938$ & &$-0.181$ &$-0.149$ \\
$6$  & &$-0.285$  & $-0.253$ & &$0.682$ &$0.990$ & &$-0.161$ &$-0.141$ \\
$7$  & &$-0.249$  & $-0.226$ & &$0.743$ &$1.029$ & &$-0.144$ &$-0.133$ \\
$8$  & &$-0.222$  & $-0.204$ & &$0.792$ &$1.053$ & &$-0.131$ &$-0.126$ \\
$9$  & &$-0.199$  & $-0.185$ & &$0.822$ &$1.061$ & &$-0.121$ &$-0.120$ \\
$10$ & &$-0.181$  & $-0.170$ & &$0.846$ &$1.056$ & &$-0.111$  &$-0.114$ \\
$11$ & &$-0.166$  & $-0.156$ & &$0.863$ &$1.049$ & &$-0.103$  &$-0.109$ \\
$12$ & &$-0.153$  & $-0.145$ & &$0.873$ &$1.041$ & &$-0.096$  &$-0.104$ \\
$13$ & &$-0.142$  & $-0.135$ & &$0.881$ &$1.033$ & &$-0.092$  &$-0.099$ \\
$14$ & &$-0.133$  & $-0.126$ & &$0.887$ &$1.024$ & &$-0.089$  &$-0.095$ \\
$15$ & &$-0.124$  & $-0.119$ & &$0.889$ &$1.015$ & &$-0.085$  &$-0.091$ \\
$16$ & &$-0.117$  & $-0.112$ & &$0.889$ &$1.006$ & &$-0.083$  &$-0.087$ \\
 %\multicolumn{7}{c}{NS-symmetric} \\
\bottomrule
\end{tabular}
\end{center}
\vspace{-1.0\baselineskip}
\tablefoot{
For the equatorial Rossby modes, $n$ denotes the number of radial nodes for $v_\theta$ at the equator. 
For the other two types modes, at fixed $m$, there are both modes with north-south symmetric and and antisymmetric $z$-vorticity $\zeta_{z}$ where $z$ denotes the rotational axis. 
These different dispersion relations and their connections are plotted in  Fig.~\ref{fig:disp_mixed}. 
}
 \end{table*}
%%%%%%%%%%%%%%%%%%%%%%%%end table%%%%%%%%%%%%%%%%%%%%%%%%%%%%

%___________________________________________________________
   \begin{figure*}[h]
     \centering
     \includegraphics[width=0.96\linewidth]{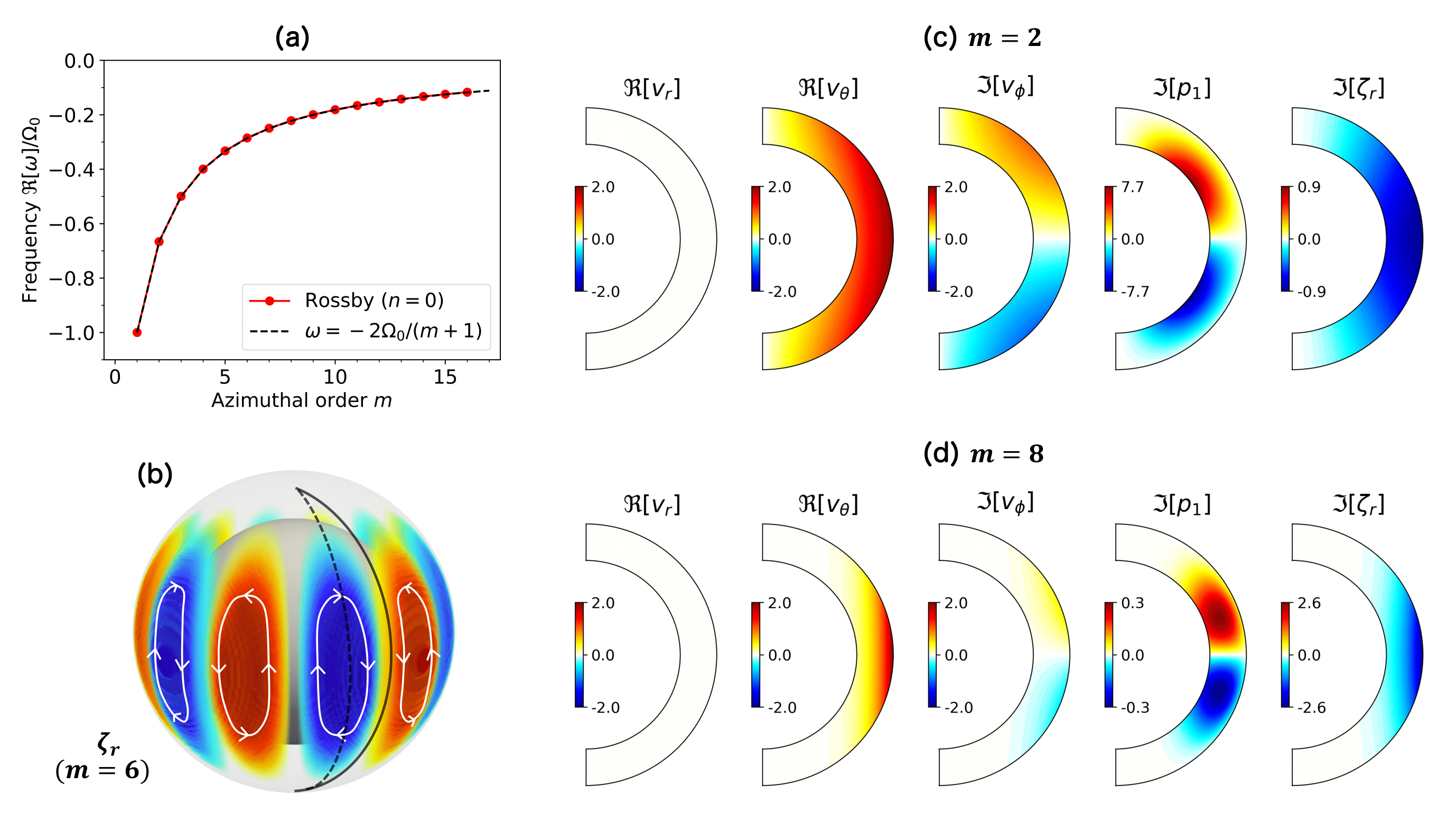}
     \caption{
     Dispersion relation and eigenfunctions of the equatorial Rossby modes without radial nodes in the case of uniform rotation, no viscosity, and adiabatic stratification.
     (a) Dispersion relation from the calculated modes (red).
     Overplotted black dashed line represents the theoretical dispersion relation of the sectoral ($l=m$) Rossby modes $\omega=2\Omega_{0}/(m+1)$.
     (b) Schematic illustration of flow structure of the mode with m=6.
     The red and blue volume rendering shows the structure of $\Re [\zeta_{r}(r,\theta) \exp{(\ii m\phi-\ii \omega t})]$.
     The black solid curve shows the meridional plane at $\phi=0$ and at $t=0$ where $v_r$ and $v_\theta$ are purely real and $v_\phi$, $p_{1}$ and $\zeta_{r}$ are purely imaginary. 
     The black dashed line denotes the meridional plane at $\phi=-\pi/2m$ where $v_\phi$, $p_{1}$ and $\zeta_{r}$ are real.
     (c) Meridional cuts of the $m=2$ eigenfunctions for the velocity
     ${\bm{v} (r,\theta)} \exp{[ \ii (m \phi -\omega t)]}$, the pressure $p_{1}(r,\theta) \exp{[\ii (m \phi -\omega t)]}$, and the  radial vorticity $\zeta_r(r,\theta) \exp{[\ii (m \phi -\omega t)]}$. 
     The solutions are shown in the meridional plane at $\phi=0$ and $t=0$.  
     The units of the color bars are m~s$^{-1}$ for the three velocity components,  $10^{5}$ dyn cm$^{-2}$ for the pressure, and $10^{-8}$ s$^{-1}$ for the
     vorticity. 
     The eigenfunctions are normalized such that the maximum of $|v_{\theta}|$ is 2 m~s$^{-1}$.
     (d) The same as panel (c) but for $m=8$.
     }
     \label{fig:eigen_eqRos}
   \end{figure*}
%___________________________________________________________
%___________________________________________________________
   \begin{figure*}[h]
     \centering
     \includegraphics[width=0.9\linewidth]{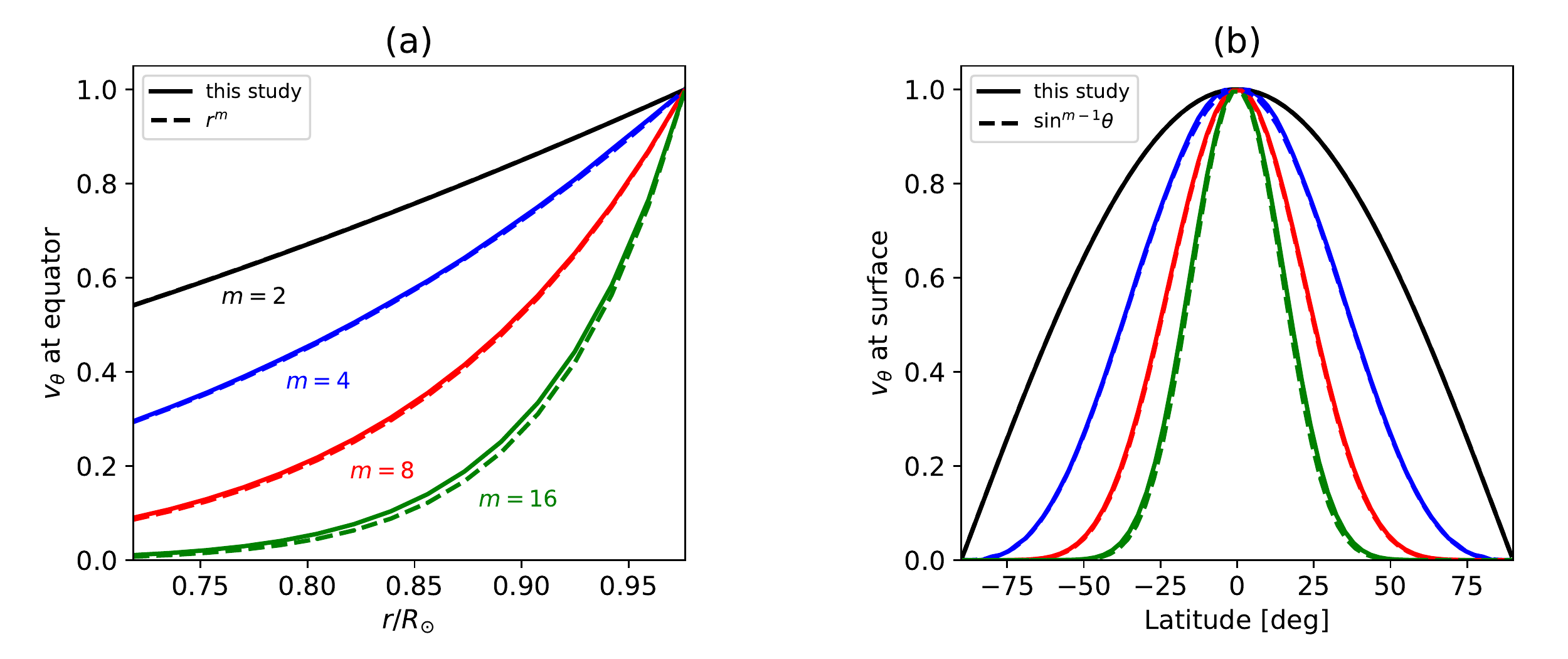}
     \caption{(a) Radial structure of the eigenfunction of $v_{\theta}$ at the equator for the $n=0$ equatorial Rossby modes in the inviscid, uniformly rotating, and adiabatically stratified case. 
     Overplotted dashed lines represent theoretically-predicted radial dependence $v_{\theta}\propto r^{m}$.
     The eigenfunctions are normalized to unity at the surface $r=r_{\mathrm{max}}$.
     (b) Latitudinal structure of the eigenfunction of $v_{\theta}$ at the surface. 
     Dashed lines are the theoretical solution in the form of legendre-polynomials $v_{\theta}\propto \sin^{m-1}{\theta}$.
     All the eigenfunctions are normalized at the equator.}
     \label{fig:vq1d_eqRos}
   \end{figure*}
%___________________________________________________________

%__________________________________________________________________
\subsection{Example spectrum for uniform rotation} \label{sec:overview}

For each $m$, there are $5N_{r}N_{\theta}$  eigensolutions with frequencies $\omega$ and eigenfunctions $\bm{V}$.
As an example, we show the typical distribution of the output eigenfrequencies in a complex plane for the case with $m=1$, $\delta=10^{-6}$ (weakly superadiabatic), and $\nu=\kappa=2\times 10^{12}$ cm$^{2}$ s$^{-1}$ in Fig.~\ref{fig:allfreq}.
Note that the differential rotation is not included here for simplicity; the uniform rotation rate $\Omega$ is equal to the Carrington rotation rate $\Omega_0$.

The modes belong to one of several regions in the complex eigenfrequency spectrum.
The modes seen in Fig.~\ref{fig:allfreq}a are acoustic modes (p modes) slightly damped due to the viscous and thermal diffusion. 
On this plot, the effect of rotation is not visible to the eye. 
In the rest of this paper, we focus on the low-frequency modes in the inertial frequency range.
Inertial oscillations are confined within the range $|\Re[\omega]|<2\Omega_{0}$ \citep[e.g.,][]{greenspan1968}.
Figure~\ref{fig:allfreq}b shows the spectrum of inertial modes in the complex plane.
The sectoral Rossby mode with no radial node ($n=0$) is easy to identify by comparison with the analytical frequency $\omega=-2 \Omega_{0}/(m+1)$.
Owing to the slightly superadiabatic background ($\delta>0$), we can see that some modes have positive imaginary frequencies ($\Im[\omega]>0$) at very low frequencies and thus are unstable.
These convective modes are shown in Fig.~\ref{fig:allfreq}c.

When the background is weakly subadiabatic (e.g., $\delta=-10^{-6}$), all the modes become stable ($\Im[\omega]<0$) and some inertial modes are partially mixed with gravity modes (g modes).
When $\Omega_0=0$, the modes are either purely convective modes or purely g modes depending on the sign of $\delta$ as shown in Fig.~\ref{fig:freqzoom_supomega}. The frequency of the g modes depends on $\delta$ and, depending on $\Omega_0$, can lie in the inertial range.

%__________________________________________________________________
%__________________________________________________________________

%%%%%%%%%%%%%%%%%%%%%%%%%%%%%%%%%%%%%%%%%%%%%%%%%%%%%%%
%%%%%%%%%%%%%%%%%%%%%%%%%%%%%%%%%%%%%%%%%%%%%%%%%%%%%%%
\section{Reference case: no diffusion,  adiabatic stratification, uniform rotation} \label{sec:results}

In this section, we report the results of an ideal case where turbulent viscous and thermal diffusivities are set to zero ($\nu=\kappa=0$), the background is convectively neutral ($\delta=0$), and no differential rotation is included ($\Omega(r,\theta)=\Omega_{0}$ and $\partial s_{0}/\partial\theta=0$).
We present the dispersion relations and eigenfunctions of various types of global-scale vorticity modes that might be relevant to the Sun.
We will use the results of this ideal setup as references and the effects of turbulent diffusion, non-adiabatic stratification, and differential rotation will later be compared to these reference results.

In the inviscid case with uniform rotation, $M$ is self adjoint, thus the physically-meaningful solutions must have real eigenfrequencies. 
We find about 10\% of the eigenfrequencies to have a nonzero imaginary part;
these correspond to numerical artifacts due to truncation errors, and the corresponding eigenfunctions have most of their power at high spatial frequencies.
For the solutions with purely real eigenfrequencies, the eigenfunctions of $v_r$ and $v_\theta$ have the same complex phase on each meridional plane, and those of $v_\phi$, $\rho_1$ are $90^{\circ}$ out of phase with respect to $v_r$ and $v_\theta$. 
In presenting the results in this section, we choose a meridional plane where $v_r$ and $v_\theta$ are real.

In the following sections, we conduct a mode-by-mode analysis for the equatorial Rossby modes with no radial nodes ($n=0$) and one radial node ($n=1$), columnar convective modes (thermal Rossby waves) with both north-south symmetries, and the high-latitude modes with both north-south symmetries.
Fundamental properties of these modes are summarized in Table~\ref{table:summary}.
Their dispersion relations are presented in Table~\ref{table:disp}.

%%%%%%%%%%%%%%%%%%%%%%%%%%%%%%%%%%%%%%%%%%%%%%%%%%%%%%%%%%%%%%%%%%%
%__________________________________________________________________
\subsection{Equatorial Rossby modes} \label{sec:r-mode}

In this section, we discuss the equatorial Rossby modes (r modes).
The modes with no radial nodes ($n=0$) and one radial node $(n=1)$ are reported.

%___________________________________________________________
   \begin{figure*}[h]
     \centering
     \includegraphics[width=0.96\linewidth]{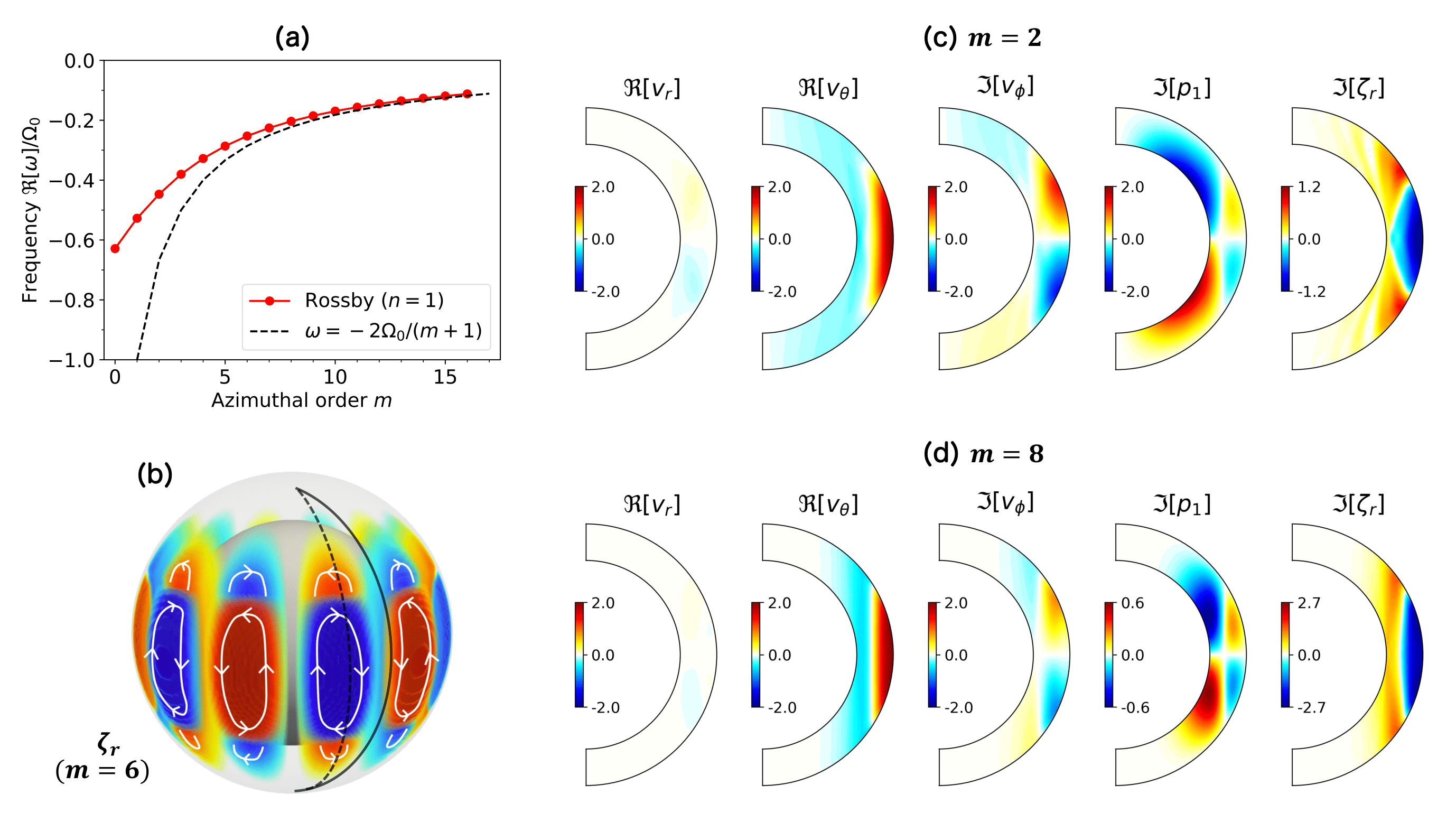}
     \caption{Dispersion relation and eigenfunctions of the equatorial Rossby modes with one radial node ($n=1$) in the inviscid, uniformly rotating, adiabatically stratified case.
     The same notation as Fig.~\ref{fig:eigen_eqRos} is used.}
     \label{fig:eigen_n1eqRos}
   \end{figure*}
%___________________________________________________________

%___________________________________________________________
   \begin{figure*}[]
     \centering
     \includegraphics[width=0.9\linewidth]{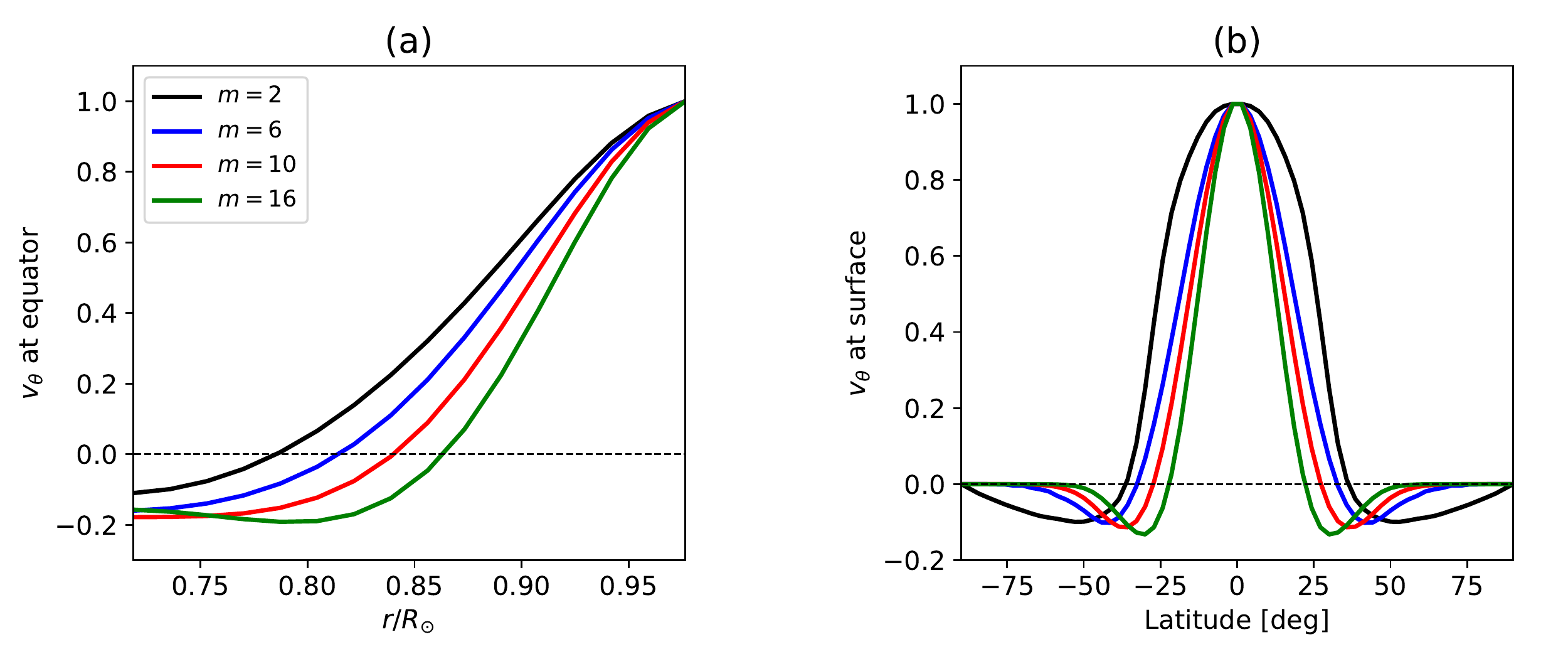}
     \caption{
     (a) Radial structure of the eigenfunction of $v_{\theta}$ at the equator of the $n=1$ equatorial Rossby modes in the inviscid, uniformly rotating, adiabatically stratified case.
      The eigenfunctions are normalized to unity at the surface $r=r_{\mathrm{max}}$.
     (b) Latitudinal structure of the eigenfunction of $v_{\theta}$ at the surface normalized at the equator.} 
     \label{fig:vq1d_n1eqRos}
   \end{figure*}
%___________________________________________________________

%___________________________________________________________
   \begin{figure*}[]
     \centering
     \includegraphics[width=0.96\linewidth]{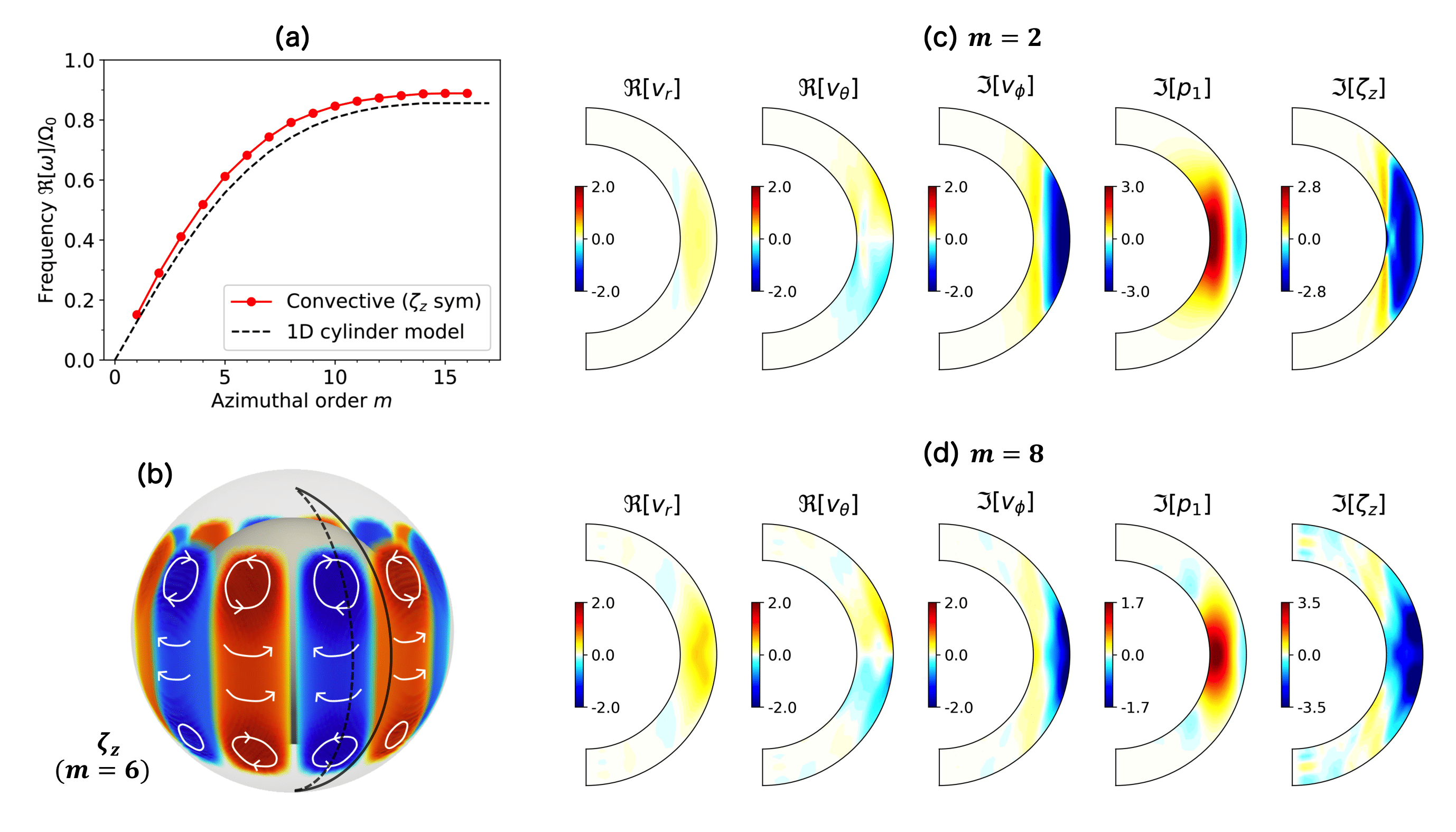}
     \caption{
     Dispersion relation and eigenfunctions of the north-south $\zeta_{z}$-symmetric columnar convective modes in the case of uniform rotation, no viscosity, and adiabatic stratification.
     (a) Dispersion relation of the north-south $\zeta_{z}$-symmetric columnar convective modes in red points.
     For comparison, dispersion relation analytically derived using one-dimensional cylinder model by \citet{glatzmaier1981} is overplotted in black dashed line.
     (b) Schematic illustration of flow structure of the mode.
     Red and blue volume rendering shows the structure of $\Re [\zeta_{z}(r,\theta) \exp{(\ii m\phi-\ii \omega t})]$ for $m=6$ at $t=0$.
     (c) Meridional cuts of the $m=2$ eigenfunctions for the velocity
     ${\bm{v} (r,\theta)} \exp{[ \ii (m \phi -\omega t)]}$
     , the pressure $p_{1}(r,\theta) \exp{[\ii (m \phi -\omega t)]}$, and the $z$-vorticity $\zeta_r(r,\theta) \exp{[\ii (m \phi -\omega t)]}$.
     The solutions are shown in the meridional plane at $\phi=0$ and $t=0$ where $v_r$ and $v_\theta$ are purely real and $v_\phi$, $p_{1}$ and $\zeta_{z}$ are purely imaginary.
     The units of velocity, pressure, and vorticity are m~s$^{-1}$, $10^{5}$ dyn cm$^{-2}$, and $10^{-8}$ s$^{-1}$, respectively.
     The eigenfunctions are normalized such that maximum of $|v_{\phi}|$ is 2 m~s$^{-1}$.
     (d) The same as panel (c) but for $m=8$.
     }
     \label{fig:eigen_thRos}
   \end{figure*}
%___________________________________________________________

%___________________________________________________________
%__________________________________________________________________
\subsubsection{$n=0$ modes} \label{sec:n=0}

In order to extract the $n=0$ equatorial Rossby mode at each $m$, we apply the following procedure to the computed eigenfunctions \bm{V}.
The latitudinal and longitudinal velocities at the surface are projected onto a  basis of associated Legendre polynomials:
\begin{eqnarray}
&& v_{\theta}(r_{\mathrm{max}},\theta)=\sum_{l=0}^{{l_{\mathrm{max}}}} a_{l-m} P_{l}^{m}(\cos{\theta}), \\
&& v_{\phi}(r_{\mathrm{max}},\theta)=\sum_{l=0}^{l_{\mathrm{max}}} b_{l-m} P_{l}^{m}(\cos{\theta}),
\end{eqnarray}
where $l_{\mathrm{max}}=2 N_\theta/3-1 = 47$. 
We also compute the number of radial nodes, $n$, of $v_\theta$ at the equator.
We select the modes that satisfy all of the following three criteria:
\begin{itemize}\setlength{\itemsep}{3pt} 
    \item[$\bullet$] The $l=m$ component of $v_\theta$ is dominant  ($|a_{0}|>|a_j|$  for all $j > 0$),
    \item[$\bullet$] the $l=m+1$ component of $v_{\phi}$ is dominant  ($|b_{1}|> |b_{j}|$  for all $j \ne 1$),
    \item[$\bullet$] and the number of radial nodes of $v_\theta$ is zero  at the equator, $n=0$.
\end{itemize}

Figure~\ref{fig:eigen_eqRos}a shows the dispersion relation of the selected $n=0$ equatorial Rossby modes for this ideal setup for $m=1-16$.
It should be noted that these modes are the only type of inertial modes where a simple analytical solution can be found in the inviscid, uniformly-rotating limit \citep[e.g.,][]{saio1982}.
Therefore, we use this analytical solution to verify our code.
The red points and black dashed lines represent the computed eigenfrequencies in our model and the theoretically-expected dispersion relation, $\omega=-2\Omega_{0}/(m+1)$, respectively.
We find that the differences in the normalized frequencies are less than $10^{-2}$ at all $m$.

The typical flow structure of this mode is schematically illustrated in Fig.~\ref{fig:eigen_eqRos}b where the volume rendering of the radial vorticity $\zeta_{r}$ is shown by red and blue.
Figures~\ref{fig:eigen_eqRos}c and~d show the real eigenfunctions for $m=2$ and $8$, respectively.
The eigenfunctions are normalized such that the maximum of $v_{\theta}$ is $2$ m~s$^{-1}$ at the surface.
The amplitude of radial velocity $v_{r}$ is about $10^{3}$ times smaller than those of horizontal velocities $v_{\theta}$ and $v_{\phi}$, implying that the fluid motion is essentially toroidal.
We find that using a higher resolution leads to even smaller $v_{r}$.
The pressure perturbation $p_{1}$ is positive (negative) where the radial vorticity $\zeta_{r}$ is negative (positive) in the northern (southern) hemisphere, which is consistent with the modes being in geostrophic balance.
As $m$ increases, the $n=0$ equatorial Rossby modes are shifted to the surface and to the equator.
The horizontal eigenfunction of $\zeta_{r}$ becomes more elongated in latitude, which means that $v_{\theta}$ becomes much stronger than $v_{\phi}$ to keep the mass conservation horizontally.

Figure~\ref{fig:vq1d_eqRos}a shows the radial structure of the  eigenfunctions of $v_{\theta}$ at the equator for selected azimuthal orders $m$.
Solid and dashed lines compare our results with the analytical solution $v_{\theta}\propto r^{m}$.
It is seen that computed eigenfunctions exhibit the $r^{m}$ dependence that agree with the analytical solutions. 
We also confirm the same $r^{m}$ dependence for the eigenfunctions of $v_{\theta}$ in the middle latitudes (not shown).
For higher $m$, the radial eigenfunction shows a slight deviation (within a few percent error) from the analytical solution.
This is possibly due to the stress-free boundary condition, $\partial (v_{\theta}/r)/\partial r=0$, at the top and bottom boundaries, which conflicts with the $r^{m}$ dependence.
Figure~\ref{fig:vq1d_eqRos}b shows the latitudinal eigenfunctions of $v_{\theta}$ at the surface.
Again, an agreement can be seen between our results and the analytical solutions $v_{\theta}\propto \sin^{m-1}{\theta}$.

%___________________________________________________________
   \begin{figure*}
     \centering
     \includegraphics[width=0.96\linewidth]{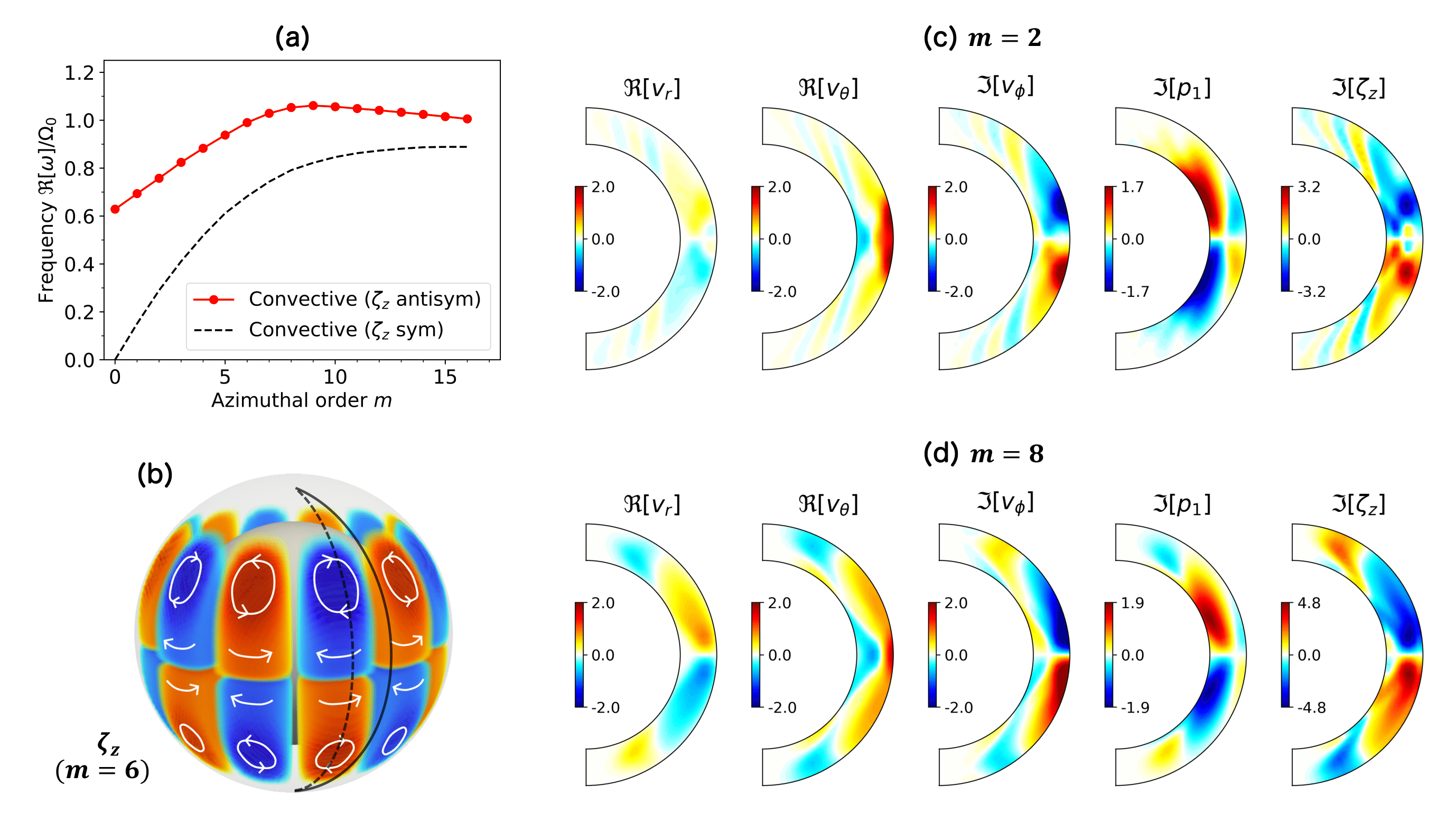}
     \caption{
     Dispersion relation and eigenfunctions of the north-south $\zeta_z$-antisymmetric columnar convective modes in the inviscid, uniformly rotating, adiabatically stratified case.
     The same notation as Fig.~\ref{fig:eigen_thRos} is used.
     In panel (a), the dispersion relation of the north-south $\zeta_z$-symmetric columnar convective modes is shown in black dashed line for comparison.
     }
     \label{fig:eigen_asthRos}
   \end{figure*}
%___________________________________________________________

%___________________________________________________________
   \begin{figure}[h]
    \centering
     \includegraphics[width=0.98\linewidth]{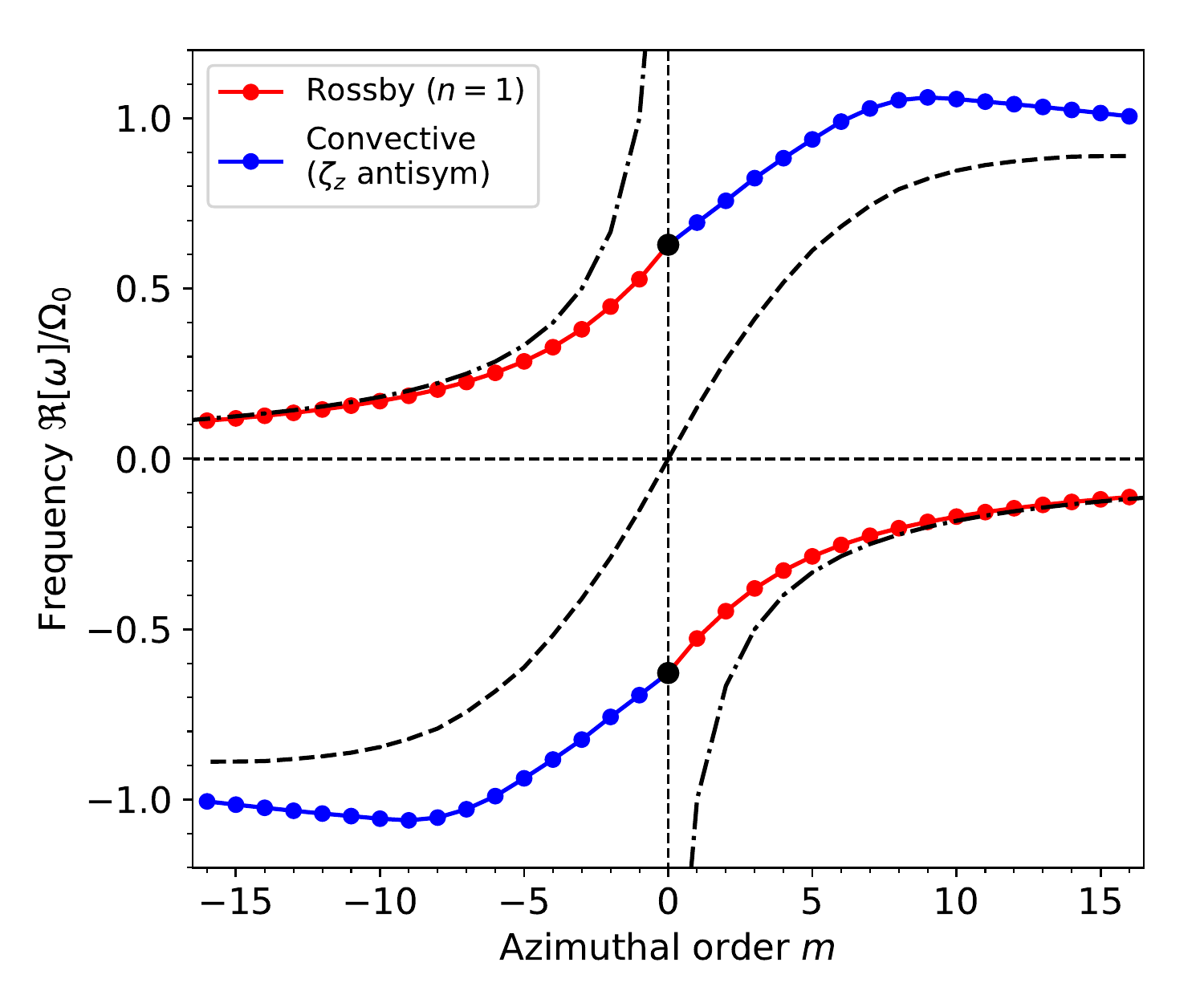}
     \caption{
     Dispersion relation of the ``mixed modes'' between the $n=1$ equatorial Rossby modes (red) and the north-south $\zeta_z$-antisymmetric columnar convective modes (blue) in the inviscid, uniformly rotating, adiabatically stratified case.
     The black points denote the axisymmetric mode at $m=0$.
     Black solid dashed and dot-dashed lines represent the dispersion relation of the $n=0$ equatorial Rossby modes and north-south $\zeta_z$-symmetric columnar convective modes.
     }
     \label{fig:disp_mixed}
   \end{figure}
%___________________________________________________________

%___________________________________________________________
   \begin{figure*}
     \centering
     \includegraphics[width=0.96\linewidth]{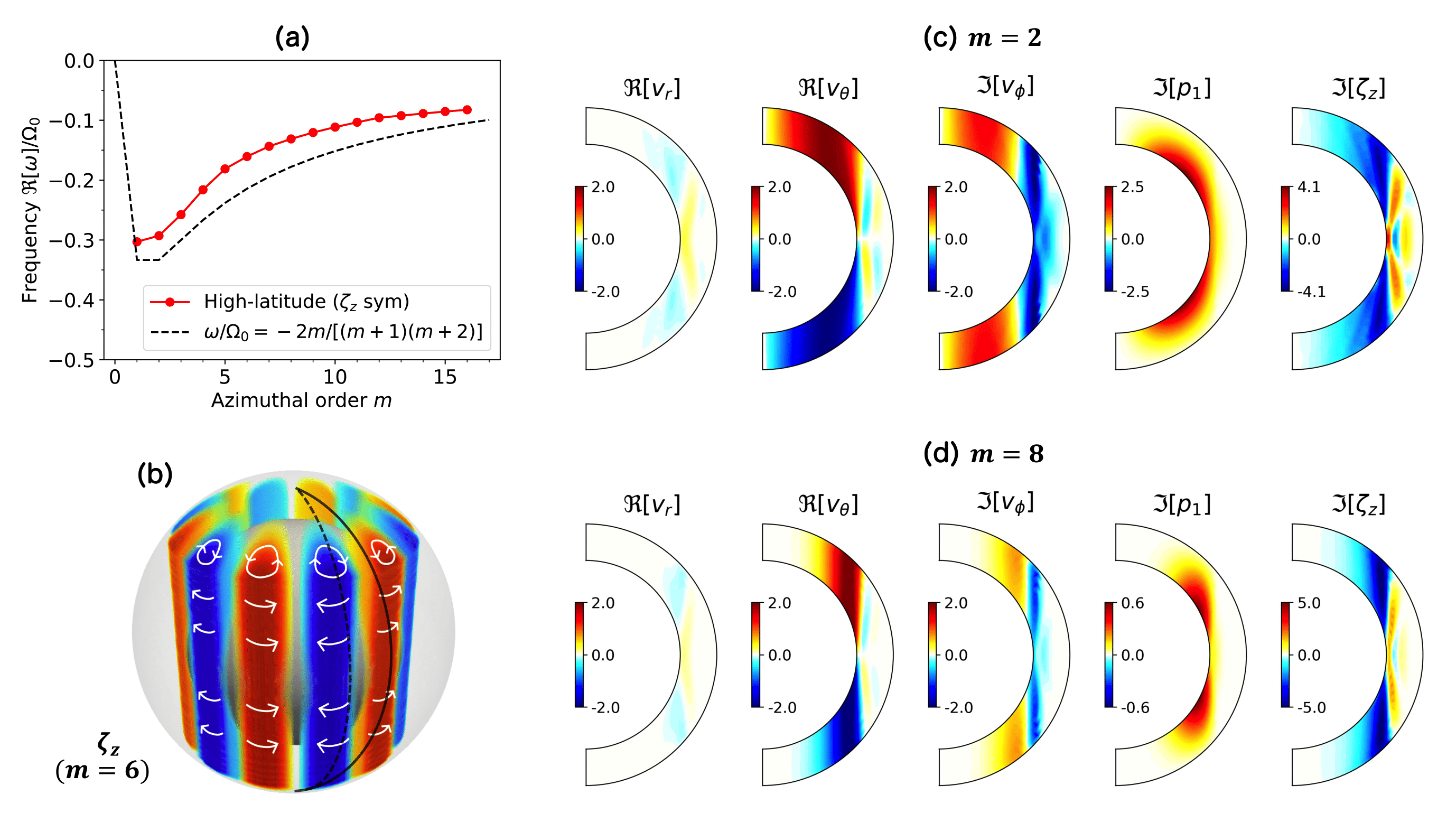}
     \caption{
     Dispersion relation and eigenfunctions of the high-latitude modes with north-south symmetric $\zeta_z$ in the inviscid, uniformly rotating, adiabatically stratified case.
     The same notation as Fig.~\ref{fig:eigen_thRos} is used.
     In panel (a), the dispersion relation of the $l=m+1$ Rossby modes is shown in black dashed line.
     }
     \label{fig:eigen_stpgRos}
   \end{figure*}
%___________________________________________________________
%___________________________________________________________
   \begin{figure*}[h]
     \centering
     \includegraphics[width=0.96\linewidth]{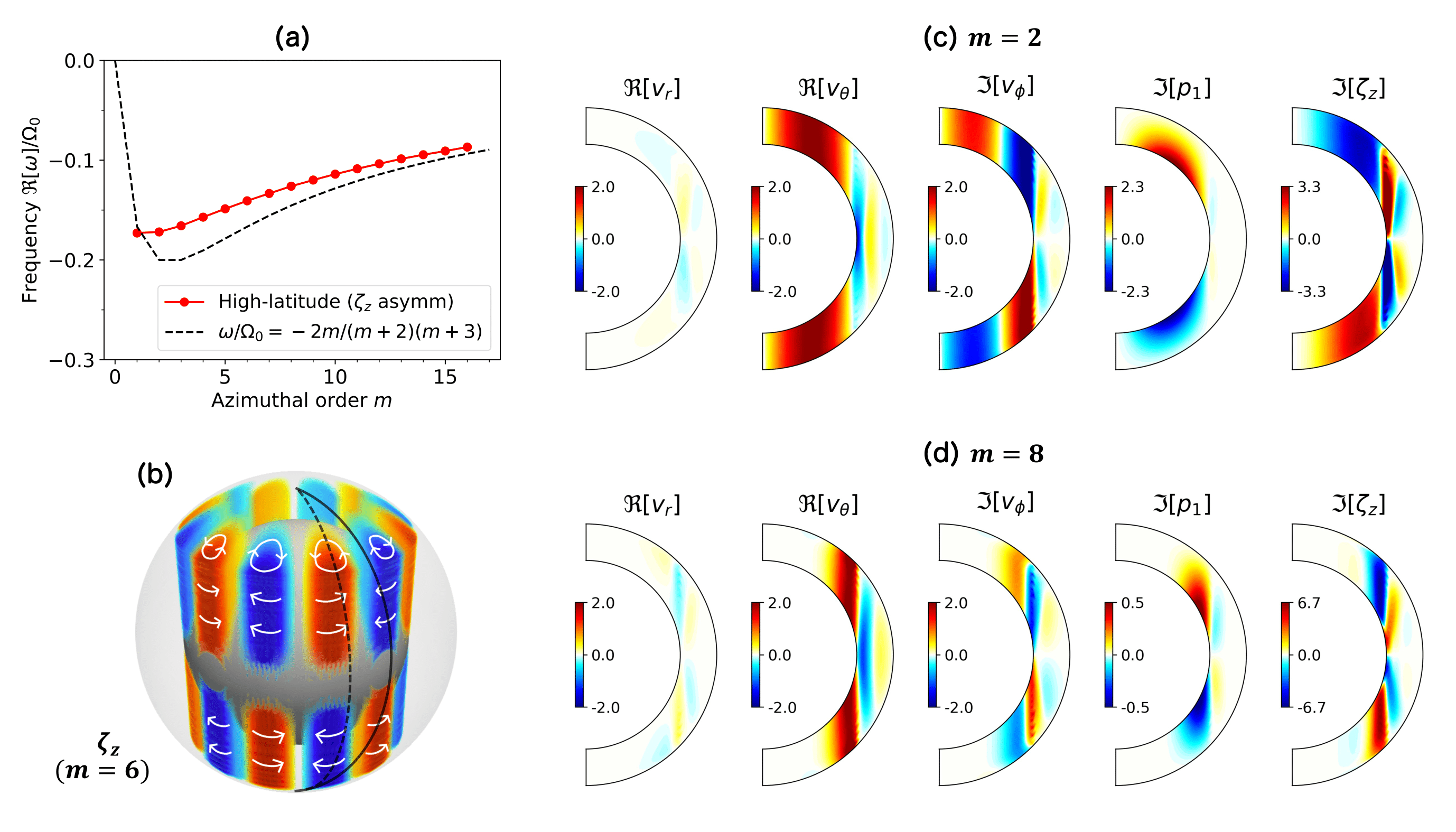}
     \caption{
     Dispersion relation and eigenfunctions of north-south $\zeta_z$-antisymmetric high-latitude modes in the inviscid, uniformly rotating, adiabatically stratified case.
     The same notation as Fig.~\ref{fig:eigen_stpgRos} is used.
     In panel (a), the dispersion relation of the $l=m+2$ Rossby mode is shown in black dashed line.
     }
     \label{fig:eigen_astpgRos}
   \end{figure*}
%___________________________________________________________

%%%%%%%%%%%%%%%%%%%%%%%%%%%%%%%%%%%%%%%%%%%
%__________________________________________________________________
\subsubsection{$n=1$ modes} \label{sec:n=1}

The equatorial Rossby modes with one radial node ($n=1$) can be selected by applying the following filters for latitudinal and longitudinal velocity eigenfunctions:
\begin{itemize}\setlength{\itemsep}{3pt} 
    \item[$\bullet$] The $l=m$ component of $v_{\theta}$ is dominant at the surface,
    \item[$\bullet$] the $l=m+1$ component of $v_{\phi}$ is dominant at the surface,
    \item[$\bullet$] and the number of radial nodes of $v_\theta$ is one at the equator.
\end{itemize}

Figure~\ref{fig:eigen_n1eqRos}a shows the dispersion relation of the selected $n=1$ equatorial Rossby modes for $0 \leq m \leq 16$.
It should be noted that we successfully identify the axisymmetric mode ($m=0$) at $\Re[\omega]=-0.63\Omega_{0}$.
This $m=0$ mode is an equatorially-trapped axisymmetric inertial mode.
It will be shown later in \S\ref{sec:thNS-AS} that this mode is connected to a prograde-propagating columnar convective mode.
The $n=1$ Rossby modes propagate in a retrograde direction with slower phase speed than that of $n=0$ Rossby modes at low $m$.
However, for $m \geq 8$, the mode frequencies become so close to those of $n=0$ modes that they are almost indistinguishable.

Figure~\ref{fig:eigen_n1eqRos}b shows a schematic sketch of typical flow motion of the $n=1$ equatorial Rossby mode.
Figures~\ref{fig:eigen_n1eqRos}c and d further shows the obtained eigenfunctions of $n=1$ equatorial Rossby modes plotted in the same way as in Fig.~\ref{fig:eigen_eqRos}.
It is clearly shown that $v_{\theta}$ has a nodal plane in the middle convection zone at the equator which extends in the direction of the rotation axis.
One of the most striking consequences of the existence of the radial node is that substantial $v_{r}$ is involved owing to the radial shear of $v_{\theta}$.
Therefore, unlike the $n=0$ modes, the associated fluid motions are no longer purely toroidal and become essentially three-dimensional.

Figure~\ref{fig:vq1d_n1eqRos}a shows the radial structure of the eigenfunctions of $v_{\theta}$ at the equator for selected $m$.
It is clearly seen that the location of the radial node shifts towards the surface as $m$ increases.
Figure~\ref{fig:vq1d_n1eqRos}b shows the latitudinal structure of the eigenfunctions of $v_{\theta}$ at the surface.
The eigenfunctions peak at the equator and change their sign in the middle latitudes ($25^{\circ}-50^{\circ}$) and decay at higher latitudes.

%%%%%%%%%%%%%%%%%%%%%%%%%%%%%%%%%%%%%%%%%%%
%__________________________________________________________________
\subsection{Columnar Convective Modes} \label{sec:thRos}

In this section, we carry out a similar mode-by-mode analysis for the columnar convective modes (thermal Rossby waves) with both hemispheric symmetries.
Here, we define the north-south symmetry based on the eigenfunction of $z$-vorticity $\zeta_{z}$.
The ``banana cel'' convection pattern can be essentially regarded as the north-south symmetric part of these convective modes.
We will also show that the north-south $\zeta_z$-antisymmetric modes are essentially mixed with the $n=1$ equatorial modes.

%%%%%%%%%%%%%%%%%%%%%%%%%%%%%%%%%%%%%%%%%%%
%__________________________________________________________________
\subsubsection{North-south $\zeta_z$-symmetric modes} \label{sec:thNS-S}

North-south $\zeta_z$-symmetric columnar convective modes can be selected by applying the following filters on the velocity eigenfunctions:
\begin{itemize}\setlength{\itemsep}{3pt} 
    \item[$\bullet$] The $l=m$ component of $v_{\phi}$ is dominant at the surface,
    \item[$\bullet$] the $l=m+1$ component of $v_{\theta}$ is dominant at the surface.
    \item[$\bullet$] the number of radial nodes of $v_{r}$ is zero at the equator,
    \item[$\bullet$] and the number of radial nodes of $v_{\phi}$ is one at the equator.
\end{itemize}

Figure~\ref{fig:eigen_thRos}a shows the dispersion relation of the selected north-south $\zeta_z$-symmetric columnar convective modes.
For comparison, we overplot in black dashed line the dispersion relation derived from the one-dimensional cylinder model of \citet{glatzmaier1981} (their figure~2).
Qualitatively, they both show similar features:
Columnar convective modes propagate in a prograde direction at all $m$.
The modes are almost non-dispersive at low $m$ ($\leq7$), but at higher $m$, the mode frequencies become almost constant at $\Re[\omega]\approx 0.85 \Omega_{0}$.
Quantitatively, our model produces the mode frequencies slightly higher (less than $10\%$) than that of the one-dimensional cylinder model.
This difference likely comes from the spherical geometry of our model:
Our model takes into account both compressional and topographic $\beta$-effects that both lead to a prograde phase propagation, whereas only compressional $\beta$-effect is included in the cylinder model of \citet{glatzmaier1981}.

Figures~\ref{fig:eigen_thRos}c and d show example eigenfunctions of the north-south $\zeta_z$-symmetric columnar convective modes.
The flow structure is dominantly characterized by the longitudinal velocity shear outside the tangential cylinder, leading to a strong $z$-vorticity (where $z$ is a coordinate in the direction of the rotation axis).
Substantial radial motions are involved where $v_{\phi}$ converges or diverges in longitudes, as schematically illustrated in Fig.~\ref{fig:eigen_thRos}b. 
Owing to the spherical curvature of the top boundary, equatorward (poleward) latitudinal flows are involved where radial flows are outward (inward).
The $z$-vortex tubes outside the tangential cylinder are often called as Taylor columns or Busse columns in the geophysical context \citep{busse1970,busse2002} or Banana cells in the solar context \citep{miesch2000}.
The pressure perturbation $p_{1}$ is generally positive (negative) where $z$-vorticity $\zeta_{z}$ is negative (positive), as the modes are in geostrophic balance.
As $m$ increases, the modes are more concentrated towards the surface and towards the equator.

%%%%%%%%%%%%%%%%%%%%%%%%%%%%%%%%%%%%%%%%%%%
%__________________________________________________________________
\subsubsection{North-south $\zeta_z$-antisymmetric modes} \label{sec:thNS-AS}

North-south $\zeta_z$-antisymmetric columnar convective modes can be selected by filtering out the eigenfunctions that satisfy the followings:
\begin{itemize}\setlength{\itemsep}{3pt} 
    \item[$\bullet$] The $l=m$ component of $v_{\theta}$ is dominant at the surface,
    \item[$\bullet$] the $l=m+1$ component of $v_{\phi}$ is dominant at the surface,
    \item[$\bullet$] and the number of radial nodes of $v_{\theta}$ is one at the equator.
\end{itemize}

The dispersion relation of the $\zeta_z$-antisymmetric columnar convective modes is shown in Fig.~\ref{fig:eigen_asthRos}a.
For comparison, we also show the dispersion relation of the $\zeta_z$-symmetric modes in black dashed line.
The modes propagate in a prograde direction with faster phase speed than that of the $\zeta_z$-symmetric modes.
At high $m$, the dispersion relation asymptotically approaches that of the $\zeta_z$-symmetric modes.

Figures~\ref{fig:eigen_asthRos}c and d show the example eigenfunctions of the north-south $\zeta_z$-antisymmetric columnar convective modes.
The flow structure is dominantly characterized by $z$-vortex tubes that are antisymmetric across the equator.
It should be noted that strong latitudinal motions are involved at the equator at the surface.

We find that the eigenfunctions of the $m=0$ mode are the complex conjugate of the $n=1$ equatorial Rossby mode, which means that these two modes are identical at $m=0$ (note the phase speed does no longer matter for the non-propagating axisymmetric mode).
To better illustrate this point, we show in Fig.~\ref{fig:disp_mixed} the dispersion relations of these two modes in the full ($m,\Re[\omega]$) domain extended to negative azimuthal orders.
It is seen that the dispersion relations of these two modes connects across $m=0$ and form a single continuous curve.
This implies that these two modes are essentially mixed with each other:
The $n=1$ equatorial Rossby modes and the north-south $\zeta_z$-antisymmetric columnar convective modes should be regarded as retrograde and prograde branches of the ``mixed'' (Rossby) modes.
It is instructive to note that this mode mixing can be understood as analogous to the so-called Yanai waves which are mixed modes between retrograde-propagating Rossby modes and prograde-propagating inertial-gravity modes \citep{matsuno1966,vallis2006}.

The flow structure itself of the $\zeta_z$-antisymmetric columnar convective mode has been recognized to be convectively-unstable in the previous literature \citep{Lorenzani2001,tilgner2007}.
However, its relation to the $n=1$ equatorial Rossby modes has never been reported.

%___________________________________________________________
   \begin{figure}
     \centering
     \includegraphics[width=\linewidth]{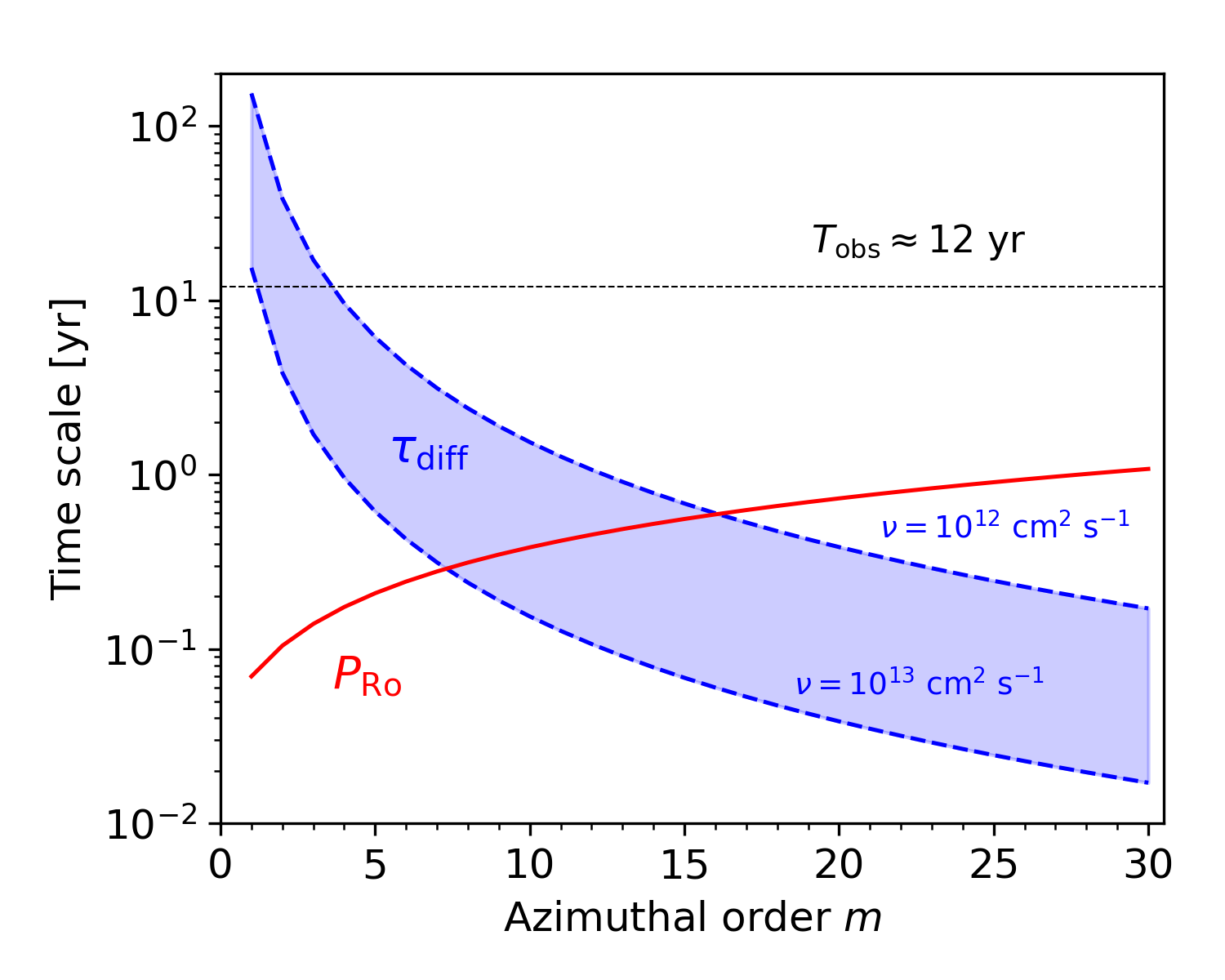}
     \caption{
     Comparison between the oscillation periods of Rossby modes $P_{\mathrm{Ro}}$ and the diffusive time scale $\tau_{\mathrm{diff}}$ for two representative values of turbulent diffusivities $\nu=10^{12}$ and $10^{13}$ cm$^{2}$~s$^{-1}$.
     The horizontal black dashed line represents the length of the SDO/HMI observational record $T_{\mathrm{obs}}\approx 12$ years.
     }
     \label{fig:mcrit}
   \end{figure}
%___________________________________________________________

%%%%%%%%%%%%%%%%%%%%%%%%%%%%%%%%%%%%%%%%%%%
%__________________________________________________________________
\subsection{High latitude modes} \label{sec:tpgRos}

In this subsection, we present the eigenmodes of the high-latitude inertial modes with both hemispheric symmetries.

%%%%%%%%%%%%%%%%%%%%%%%%%%%%%%%%%%%%%%%%%%%
%__________________________________________________________________
\subsubsection{North-south $\zeta_z$-symmetric modes} \label{sec:tpgNS-S}

To discuss these modes, it is useful to introduce a cylindrical coordinate system $(\varpi,\phi,z)$. 
In this coordinate system, the tangent cylinder is located at $\varpi=r_{\mathrm{min}}$, i.e., it is the cylinder aligned with the rotation axis which touches the radiative interior at the equator.
North-south $\zeta_z$-symmetric high-latitude modes can be selected by applying the following criteria: 
\begin{itemize}\setlength{\itemsep}{3pt} 
    \item[$\bullet$] The kinetic energy is predominantly inside the tangential cylinder, i.e., $E_{\mathrm{in}}/E_{\mathrm{CZ}}> 0.5$ where $E_{\mathrm{in}}$ and $E_{\mathrm{CZ}}$ are the volume-integrated kinetic energies inside the tangent cylinder and in the entire convection zone, respectively.
     \item[$\bullet$] The  $l=m+1$ component of $v_{\theta}$ is dominant at the bottom of the convection zone.	
    \item[$\bullet$] The number of $z$-nodes of $v_{\theta}$ is zero at $\varpi=0.5R_{\odot}$. 
\end{itemize}

Figure~\ref{fig:eigen_stpgRos}a shows the dispersion relation of the north-south $\zeta_z$-symmetric high-latitude modes.
We find the high-latitude modes are much more dispersive than the columnar convective modes at low $m$.
The dispersion relation is found to be roughly approximated by the non-sectoral Rossby modes' dispersion relation with one latitudinal node ($l=m+1$), as shown in the black dashed line in Fig.~\ref{fig:eigen_stpgRos}a.
This is because the horizontal flows at the bottom boundary behave like the $l=m+1$ (classical) Rossby modes.
Note, however, that this is not regarded as the mode mixing as discussed in \S\ref{sec:thNS-AS}.
%The genuine $l=m+1$ Rossby modes are identified to exist near the surface equatorial region (not shown).

Figures~\ref{fig:eigen_stpgRos}c and d show example eigenfunctions of the $\zeta_z$-symmetric high-latitude modes.
The fluid motion is predominantly characterized by $z$-vortices inside the tangential cylinder in both hemispheres, as schematically illustrated in the Fig.~\ref{fig:eigen_stpgRos}b.
The power of $\zeta_{z}$ peaks at the tangential cylinder $\varpi=r_{\mathrm{min}}$.
Note that the longitudinal velocity $v_{\phi}$ extends slightly outside the cylinder.
Again, $\Im[p_{1}]\Im[\zeta_{z}]<0$ follows from the mode being in geostrophic balance.

%___________________________________________________________
   \begin{figure*}[]
     \centering
     \includegraphics[width=0.85\linewidth]{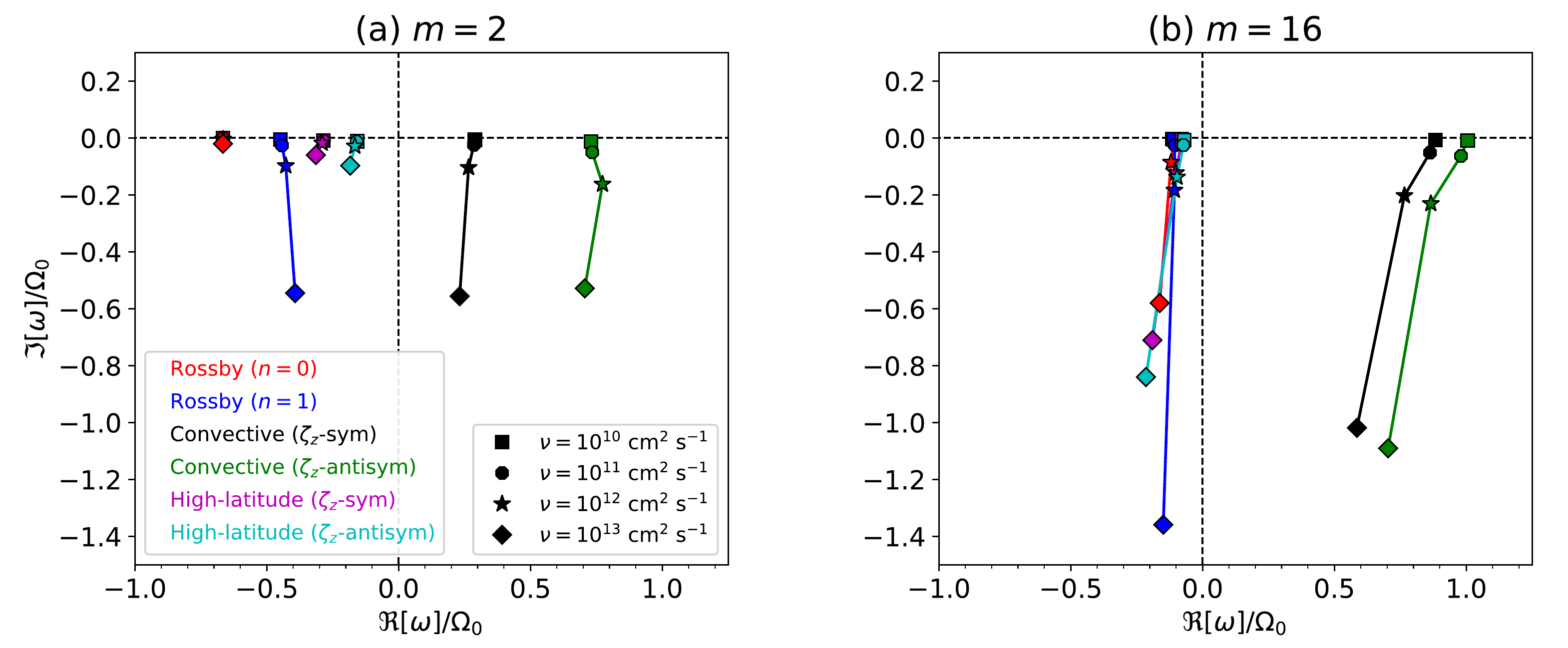}
     \caption{
     Eigenfrequency spectra of the low-frequency vorticity modes in a complex plane with different values of diffusivities for (a) $m=2$ and (b) $m=16$.
     Different colors represent different classes of inertial modes.
     Different symbols represent different values of the viscous and thermal diffusivities.
     In all cases, rotation is uniform and the stratification is adiabatic.
     }
     \label{fig:eigfreq_diff}
   \end{figure*}
%___________________________________________________________
%___________________________________________________________
   \begin{figure*}[]
     \centering
     \includegraphics[width=0.865\linewidth]{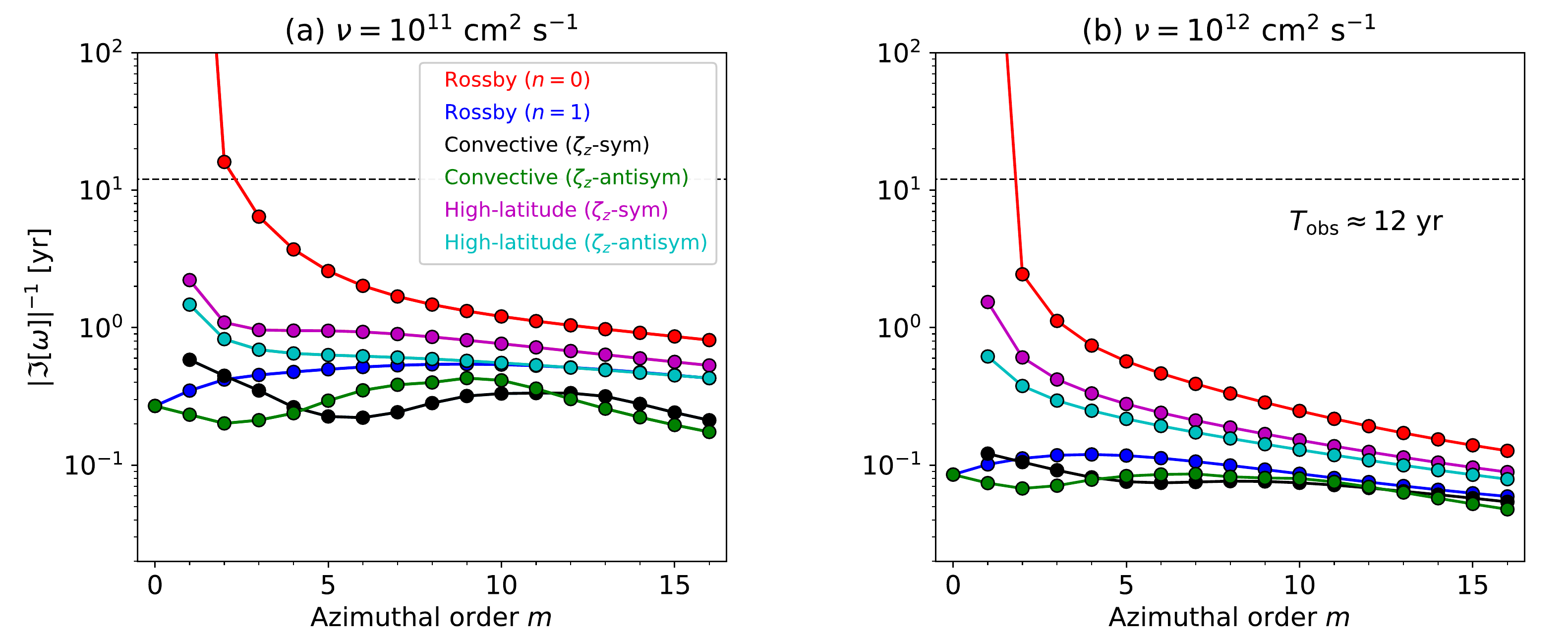}
     \caption{
     $e$-folding lifetimes of various low-frequency modes for a viscous diffusivity  (a) $\nu=10^{11}$ cm$^{2}$ s$^{-1}$ and (b) $\nu=10^{12}$ cm$^{2}$ s$^{-1}$.
     Note that all the  modes selected here are stable modes ($\Im[\omega]<0$).
     Different colors represent different types of inertial modes.
     The horizontal black dashed line shows the length of the SDO/HMI observational record ($T_{\mathrm{obs}} \approx 12$ yr as of today).
     In both cases, rotation is uniform and the stratification is adiabatic.
     The lifetimes of the convective modes and high-latitude modes are very sensitive to the radial and latitudinal entropy gradients, a point which is discussed in \S~\ref{sec:delta} and \S~\ref{sec:highlat_m1}.
     }
     \label{fig:lifetime}
   \end{figure*}
%___________________________________________________________

%%%%%%%%%%%%%%%%%%%%%%%%%%%%%%%%%%%%%%%%%%%
%____________________ ______________________________________________
\subsubsection{North-south $\zeta_z$-antisymmetric modes} \label{sec:tpgNS-AS}

North-south $\zeta_z$-antisymmetric high-latitude modes are selected using the following filters:
\begin{itemize}\setlength{\itemsep}{3pt} 
    \item[$\bullet$] The kinetic energy is predominantly inside the tangential cylinder.
     \item[$\bullet$] the $l-m=1$ (or $3$) component of $v_{\phi}$ is dominant at the bottom of the convection zone,
    \item[$\bullet$] and the number of $z$-nodes is zero for $v_{\theta}$ at $\varpi=0.5R_{\odot}$.
\end{itemize}

The example modes are presented in Fig.~\ref{fig:eigen_astpgRos}.
The eigenfunctions show very similar properties of the high-latitude modes discussed in Fig.~\ref{fig:eigen_stpgRos} except for the north-south symmetry.
It should be pointed out that there exists a latitudinal flow along the tangential cylinder and across the equator.
The dispersion relation of the $\zeta_z$-antisymmetric high-latitude modes are found to be similar to that of $l=m+2$ Rossby modes, as shown in Fig.~\ref{fig:eigen_astpgRos}a.

%__________________________________________________________________
%__________________________________________________________________

%__________________________________________________________________
%__________________________________________________________________

\section{Effect of turbulent diffusion} \label{sec:viscosity}

So far, we have discussed the results for an inviscid case.
In this section, we examine the effects of viscous and thermal diffusion arising from turbulent mixing of momentum and entropy in the Sun \citep[e.g.,][]{rudiger1989}.
Let us start our discussion by estimating the impact of the turbulent diffusion on (classical) Rossby modes.
The oscillation period of the equatorial Rossby mode at the azimuthal order $m$ is given by
\begin{eqnarray}
&& P_{\mathrm{Ro}}=\left| \frac{2\pi}{\omega_{\mathrm{Ro}}} \right|,
\ \ \ \  \mathrm{where} \ \ \ \ 
\omega_{\mathrm{Ro}}=-\frac{2\Omega_{0}}{m+1}.
\end{eqnarray}
On the other hand, typical diffusive time scale can be estimated as
\begin{eqnarray}
&& \tau_{\mathrm{diff}}=\frac{l_{m}^{2}}{\nu},
\ \ \ \  \mathrm{with} \ \ \ \ 
l_{m}=\frac{R_{\odot}}{m},
\end{eqnarray}
where $l_m$ denotes the typical length scale of the Rossby mode.
Figure~\ref{fig:mcrit} compares $P_{\mathrm{Ro}}$ and $\tau_{\mathrm{diff}}$ as functions of $m$.
Two representative values of turbulent diffusitivies in the solar convection zone $\nu=10^{12}$ and $10^{13}$ cm$^{2}$~s$^{-1}$ are shown \citep[e.g.,][]{ossendrijver2003}.
When $P_{\mathrm{Ro}} \ll \tau_{\mathrm{diff}}$, viscous diffusion is almost negligible.
However, if $P_{\mathrm{Ro}} \gtrsim \tau_{\mathrm{diff}}$, diffusion can have a dominant effect on the Rossby modes.
For a given turbulent diffusivity $\nu$, the critical azimuthal order $m_{\mathrm{crit}}$ can be defined as
\begin{eqnarray}
&&m_\mathrm{crit}=\left( \frac{R_{\odot} \Omega_{0}}{\pi \nu}\right)^{1/3}. \label{eq:mcrit}
\end{eqnarray}
The Rossby modes are dominated by diffusive effects for $m>m_{\mathrm{crit}}$.
Figure~\ref{fig:mcrit} implies that the Rossby modes in the Sun are substantially affected by the turbulent diffusion especially for $m \geq 5-6$.

In this paper, we carry out a set of calculations of uniformly-rotating adiabatic fluid with varying diffusivities; $\nu=10^{9},10^{10},10^{11},10^{12}$, and $10^{13}$ cm$^{2}$ s$^{-1}$.
For simplicity, we fix the Prandtl number to unity so that $\kappa=\nu$. 
Now, both the eigenfrequencies and eigenfunctions are complex.
Figure~\ref{fig:eigfreq_diff} shows the eigenfrequencies of the six types of Rossby modes discussed in \S~\ref{sec:results} for different viscous diffusivities in a complex plane.
Figures~\ref{fig:eigfreq_diff}a and b show the cases for $m=2$ and $16$, respectively.
In general, the modes are damped by diffusion so that the imaginary frequencies are shifted towards more negative values.
At small $m$ (e.g. $m=2$), diffusion tends to act predominantly on the columnar convective modes with both symmetries and $n=1$ equatorial Rossby modes, whereas the $n=0$ Rossby modes and the high-latitude modes remain almost unaffected.
At large $m$ (e.g. $m=16$), however, all the modes are damped to a similar degree.
Note that a strong diffusion modifies not only the imaginary part but also the real part of the mode frequencies.
The computed $e$-folding times of these modes $|\Im[\omega]|^{-1}$ are shown in Figure~\ref{fig:lifetime} for the two representative values of turbulent viscosity $\nu=10^{11}$ and $10^{12}$ cm$^{2}$~s$^{-1}$.

\begin{figure*}
     \centering
     \includegraphics[width=\linewidth]{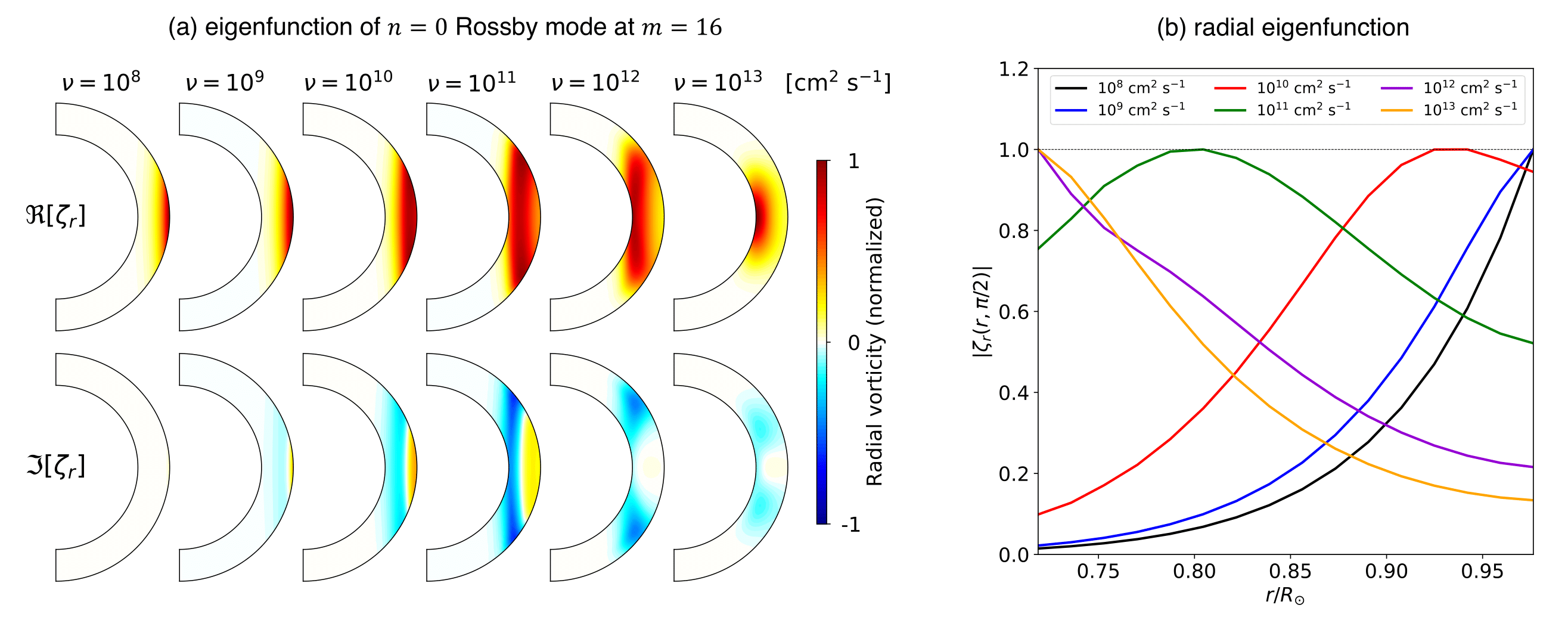}
     \caption{
     (a) Meridional eigenfunctions of radial vorticity $\zeta_{r}$ of the $n=0$ equatorial Rossby mode at $m=16$ for different values of viscous diffusivities $\nu$.
     Upper and lower panels show the normalized real and imaginary eigenfunctions, respectively.
     (b) Radial eigenfunctions of $|\zeta_{r}|$ at the equator (normalized by their maximum amplitudes).
     Different colors represent different values of diffusivities.
     In all cases, rotation is uniform and the stratification is adiabatic.
     }
     \label{fig:vort_diff}
   \end{figure*}

Now, let us focus on the $n=0$ equatorial Rossby modes to see how eigenfunctions are affected by the viscous diffusion.
Figure~\ref{fig:vort_diff}a shows the real (top row) and imaginary (bottom row) eigenfunctions of radial vorticity $\zeta_{r}$ at $m=16$ for different values of viscous diffusivities $\nu$.
As $\nu$ increases, the $n=0$ equatorial Rossby modes are shifted towards the base of the convection zone.
This is clearly illustrated in Fig.~\ref{fig:vort_diff}b where the absolute amplitudes of radial vorticity at the equator are shown as functions of radius.
When $\nu$ becomes sufficiently large, the radial eigenfunction substantially deviates from the well-known $r^{m}$ dependence. 
This can be explained as follows:
With the moderate diffusion included, the radial force balance between Coriolis force and pressure gradient force is no longer maintained.
Consequently, radial flows are driven and the diffusive momentum flux becomes directed radially inward.
In fact, the confinement of the $n=0$ equatorial Rossby modes near the base is also seen in rotating convection simulations where the diffusion can be significantly enhanced by turbulent convection (Bekki et al., in prep.).

%___________________________________________________________
%___________________________________________________________

%___________________________________________________________
   \begin{figure*}
     \centering
     \includegraphics[width=0.95\linewidth]{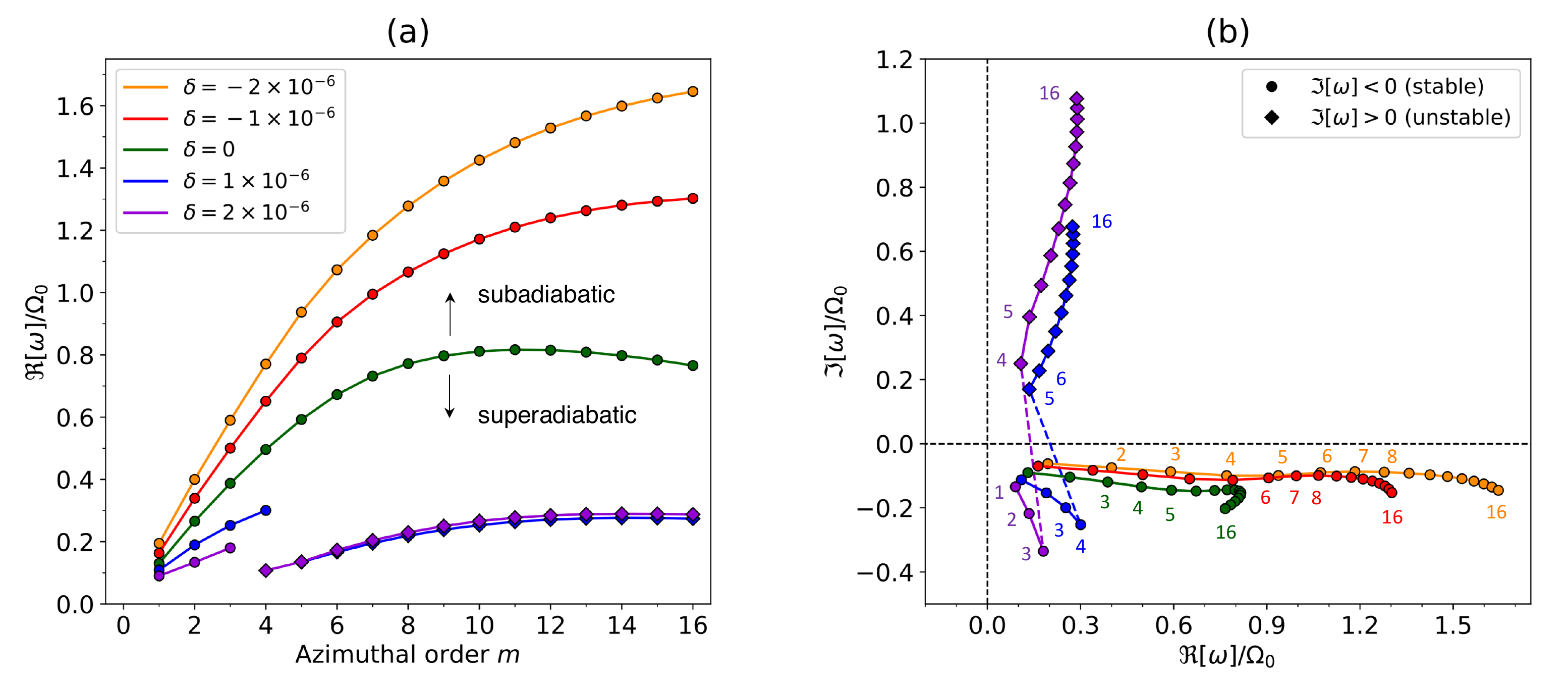}
     \caption{
     (a) Dispersion relations of the north-south $\zeta_z$-symmetric columnar convective modes with different background superadiabaticity values $\delta$.
     Different colors represent different values of superadiabaticity.
     Circles and diamonds denote the stable ($\Im[\omega]<0$) and unstable ($\Im[\omega]>0$) modes, respectively.
     (b) Eigenfrequencies in the complex plane.
     Each circle (diamond) represent a mode with azimuthal order $m$, which is labelled with small integers from $m=1$ to $16$.
     In all cases, rotation is uniform and the diffusivities are set $\nu=\kappa=10^{12}$ cm$^{2}$ s$^{-1}$.
     }
     \label{fig:eigfreq_thRos}
   \end{figure*}
%___________________________________________________________

%___________________________________________________________
   \begin{figure*}[]
     \centering
     \includegraphics[width=\linewidth]{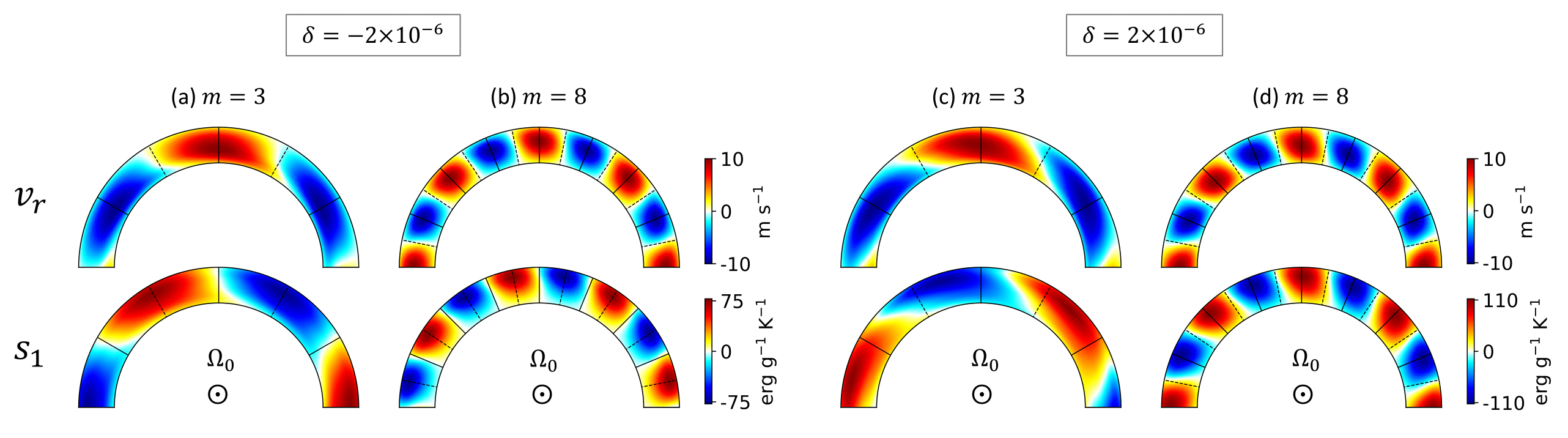}
     \caption{
     Radial velocity $v_{r}$ (upper plots) and entropy perturbation $s_{1}$ (lower plots) of the columnar convective modes along the rotational axis displayed in the equatorial plane for subadiabatic and superadiabatic background.
     Panels (a) and (b) show the cases with subadiabatic background $\delta=-2\times 10^{-6}$ for $m=3$ and $m=8$, respectively.
     Panels (c) and (d) are the same plots for superadiabatic background $\delta=2\times 10^{-6}$.
     The eigenfunctions are normalized such that the maximum radial velocity is $10$ m~s$^{-1}$ at the equator.
     In all cases, rotation is uniform and the diffusivities are set $\nu=\kappa=10^{12}$ cm$^{2}$ s$^{-1}$.
     }
     \label{fig:delta_thRos}
   \end{figure*}
%___________________________________________________________

%___________________________________________________________
   \begin{figure}%[h!]
     \centering
     \includegraphics[width=0.96\linewidth]{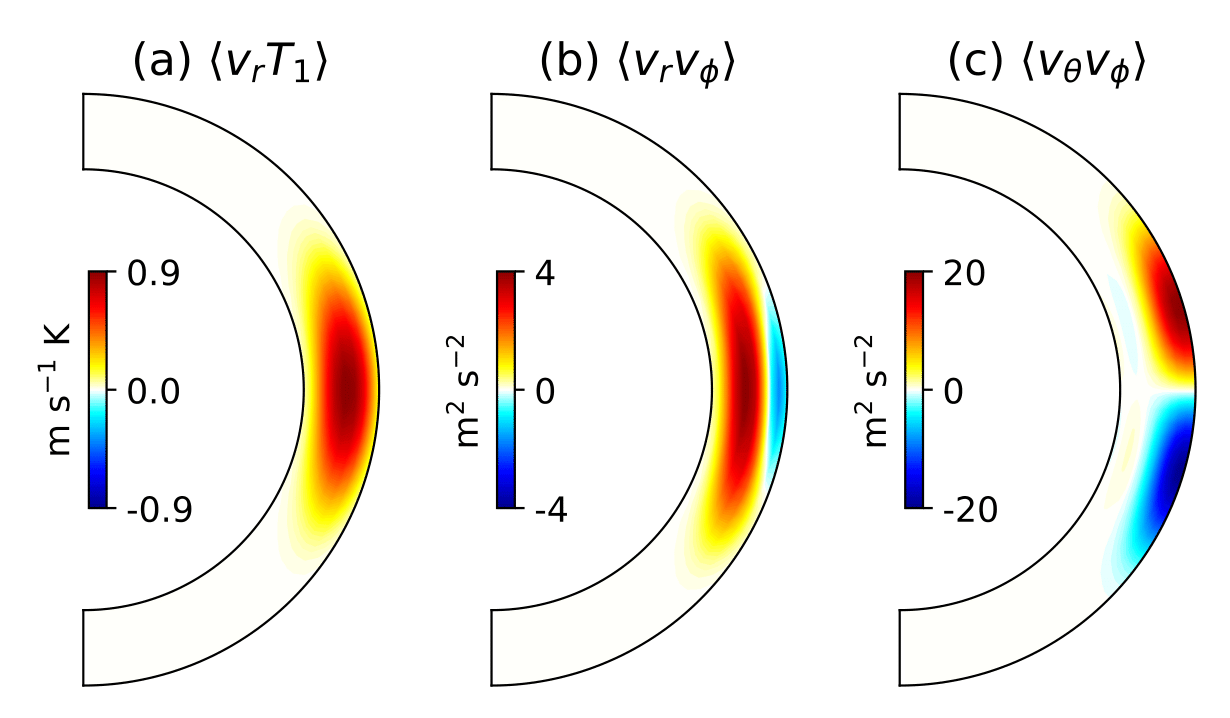}
     \caption{
     Transport properties of thermal energy and angular momentum by the north-south $\zeta_z$-symmetric columnar convective modes for $m=16$.
     (a) Correlation between radial velocity velocity and temperature perturbation $\langle v_{r}T_{1} \rangle$, (b) Reynolds stress between radial and longitudinal velocities $\langle v_{r}v_{\phi} \rangle$, and (c) Reynolds stress between latitudinal and longitudinal velocities $\langle v_{\theta}v_{\phi} \rangle$.
     The background is weakly superadiabatic ($\delta=2\times 10^{16}$), rotation is uniform, and moderate diffusivities are used ($\nu=\kappa=10^{12}$ cm$^{2}$ s$^{-1}$).
     The eigenfunctions are normalized such that the maximum radial velocity is $10$ m~s$^{-1}$ at the equator.
     }
     \label{fig:FeRS_thRos}
   \end{figure}
%___________________________________________________________

%___________________________________________________________
   \begin{figure}%[h!]
     \centering
     \includegraphics[width=0.99\linewidth]{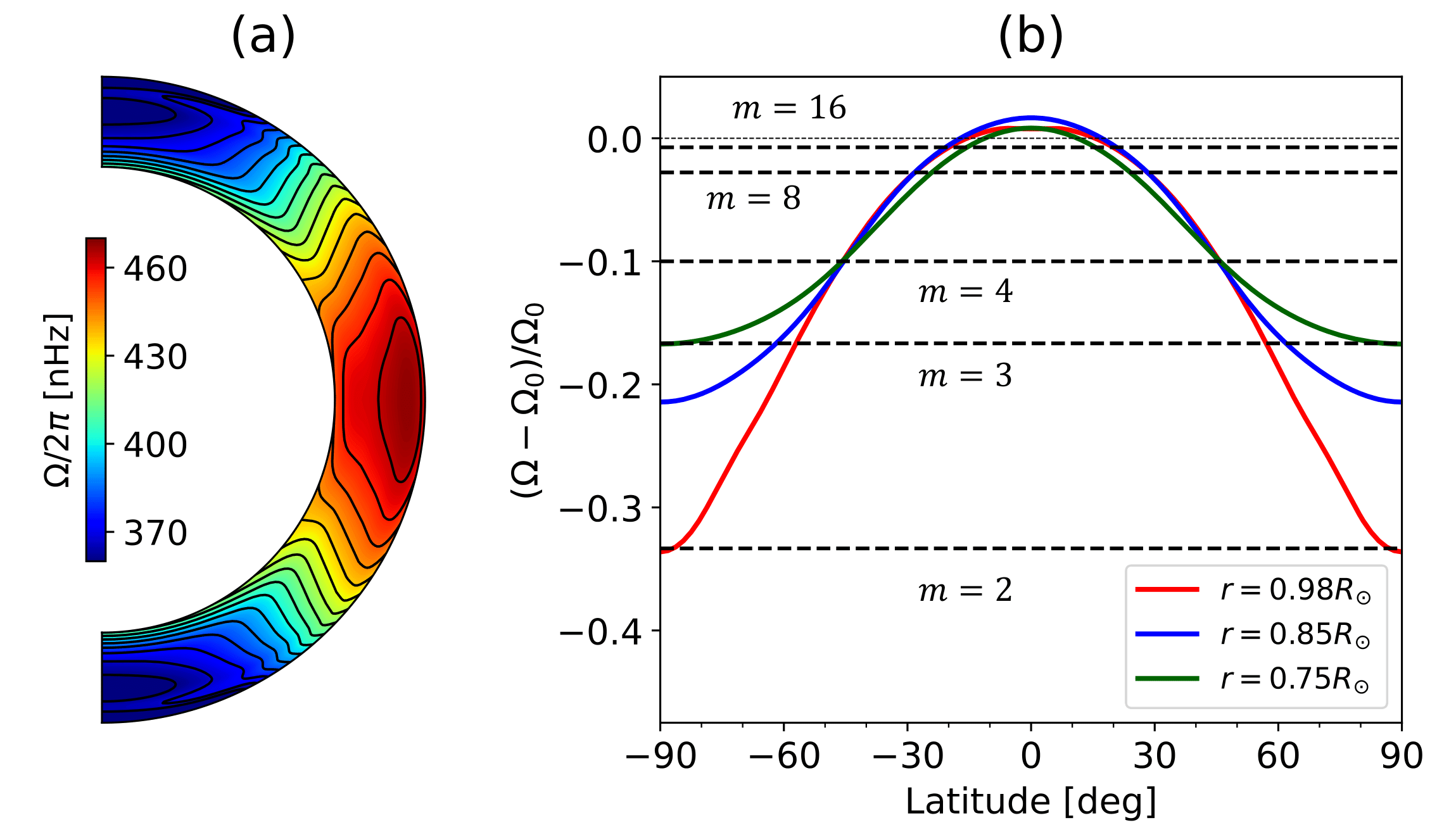}
     \caption{Solar differential rotation profile used in this study.
     (a) Differential rotation $\Omega(r,\theta)$ in a meridional plane, deduced from the global helioseismology \citep[][]{larson2018}.
     (b) Latitudinal profiles of differential rotation at different depths.
     Horizontal dashed lines indicate the theoretically-expected phase speed of the sectoral ($l=m$) classical Rossby modes for selected azimuthal orders $m=2,3,4,8,16$.
     The observing frame is chosen to be the Carrington frame rotating at $\Omega_{0}/2\pi=456.0$ nHz.
     }
     \label{fig:dr}
   \end{figure}
%___________________________________________________________

%___________________________________________________________
   \begin{figure*}[]
     \centering
     \includegraphics[width=0.85\linewidth]{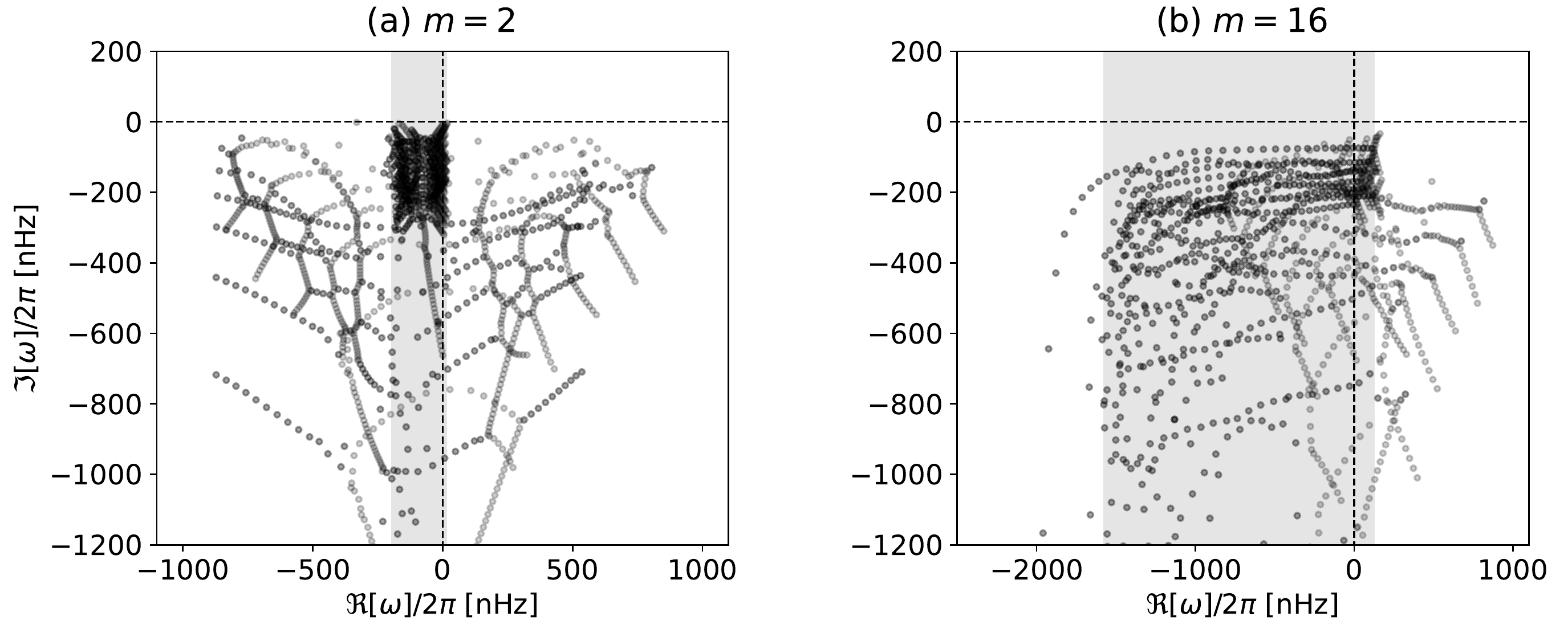}
     \caption{Eigenfrequencies of inertial modes under the solar differential rotation in the complex plane for (a) $m=2$ and (b) $m=16$.
     The diffusivity is set to $\nu=10^{12}$ cm$^{2}$ s$^{-1}$ and the background is assumed to be adiabatic, $\delta=0$.
     The shaded areas indicate the frequency range associated with the surface differential rotation, i.e., $m(\Omega_{\rm pole}-\Omega_{0})<\Re[\omega]<m(\Omega_{\rm eq}-\Omega_{0})$.
     }
     \label{fig:freq_dr}
   \end{figure*}
%___________________________________________________________

%___________________________________________________________
   \begin{figure*}[]
     \centering
     \includegraphics[width=0.91\linewidth]{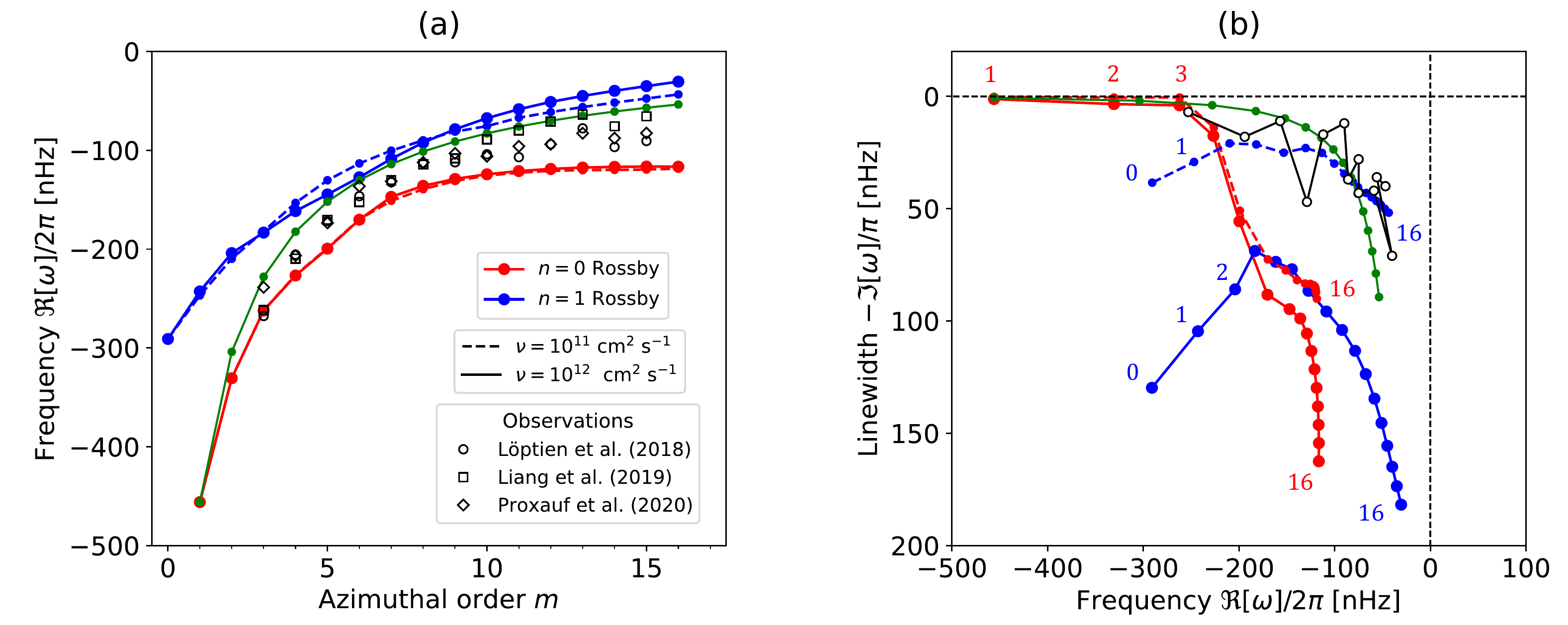}
     \caption{(a) Dispersion relations of the equatorial Rossby modes for the cases with solar differential rotation.
     Red and blue curves represent the modes with no radial nodes ($n=0$) and one radial node ($n=1$), respectively.
     Solid and dashed lines denote the cases with weak diffusion ($\nu=10^{11}$ cm$^{2}$ s$^{-1}$) and strong diffusion ($\nu=10^{12}$ cm$^{2}$ s$^{-1}$), respectively.
     For comparison, the observed Rossby mode frequencies reported in \citet{loeptien2018}, \citet{liang2019}, and \citet{proxauf2020} are plotted by white hexagons and squares.
     All the presented frequencies are measured in the Carrington frame rotating at $\Omega_{0}/2\pi=456.0$ nHz.
     (b) Mode linewidths versus mode frequencies.
     Each point represents a mode with azimuthal order $m$, which is labelled with small integers from $m=1$ to $16$.
     Overplotted (open circles connected by black lines) are the mode linewidths and frequencies measured by \citet{proxauf2020} for $m=3$ to $15$.
     In both panels, the green dots connected by line  segments refer to a theoretical model for the simplified case of uniform rotation, $\omega/\Omega_0=-2/(m+1)-\ii E_{\mathrm{k}} m(m+1)$, where $E_{\mathrm{k}}=4\times 10^{-4}$ is the Ekman number  at the solar surface \citep[][]{fournier2022}.
     }
     \label{fig:disp_eqRos_dr}
   \end{figure*}
%___________________________________________________________

%___________________________________________________________
   \begin{figure*}[]
     \centering
     \includegraphics[width=0.85\linewidth]{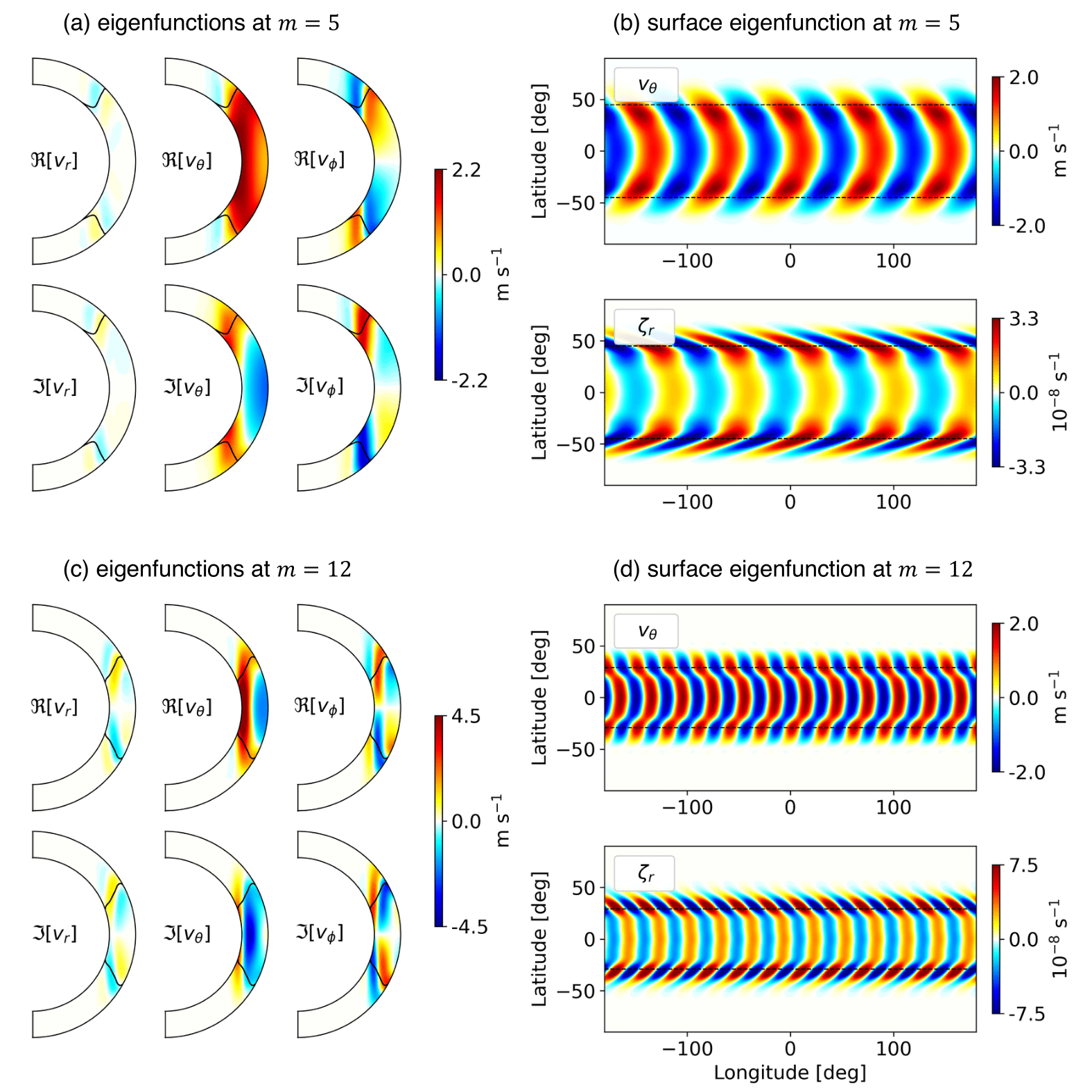}
     \caption{
     Eigenfunctions of the equatorial Rossby modes with no radial nodes ($n=0$).
     (a) Real (upper) and imaginary (lower) eigenfunctions of three components of velocity shown in a meridional plane for $m=5$.
     The eigenfunctions are normalized such that the maximum latitudinal velocity is $2$ m~s$^{-1}$ at the surface.
     The solid black line indicates the location of the critical latitudes where the phase speed of a Rossby modematches to the differential rotation sped.
     (b) Horizontal eigenfunctions of latitudinal velocity $v_{\theta}$ (upper) and radial vorticity $\zeta_{r}$ (lower) at the surface $r=0.985R_{\odot}$ for $m=5$.
     The horizontal black dashed lines indicate the location of the critical latitudes at the surface.
     (c) and (d) are the counterparts of the panels (a) and (b) for $m=12$.
     }
     \label{fig:eigenfunc_eqRos_dr}
   \end{figure*}
%___________________________________________________________

%___________________________________________________________
   \begin{figure*}[]
     \centering
     \includegraphics[width=0.85\linewidth]{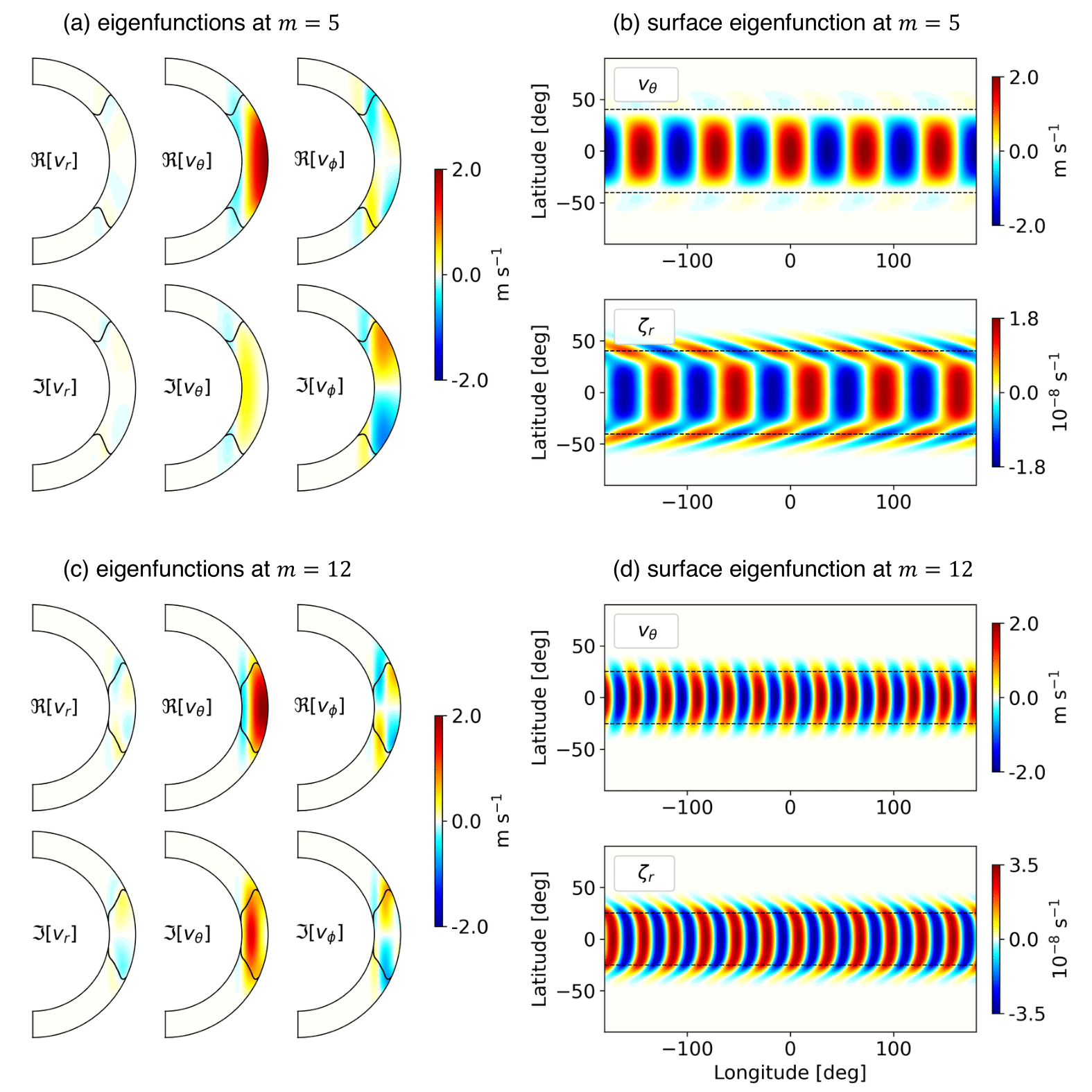}
     \caption{The same as Fig.~\ref{fig:eigenfunc_eqRos_dr} but for the equatorial Rossby modes with one radial node ($n=1$).
     }
     \label{fig:eigenfunc_n1eqRos_dr}
   \end{figure*}
%___________________________________________________________

%___________________________________________________________
   \begin{figure*}[]
     \centering
     \includegraphics[width=0.835\linewidth]{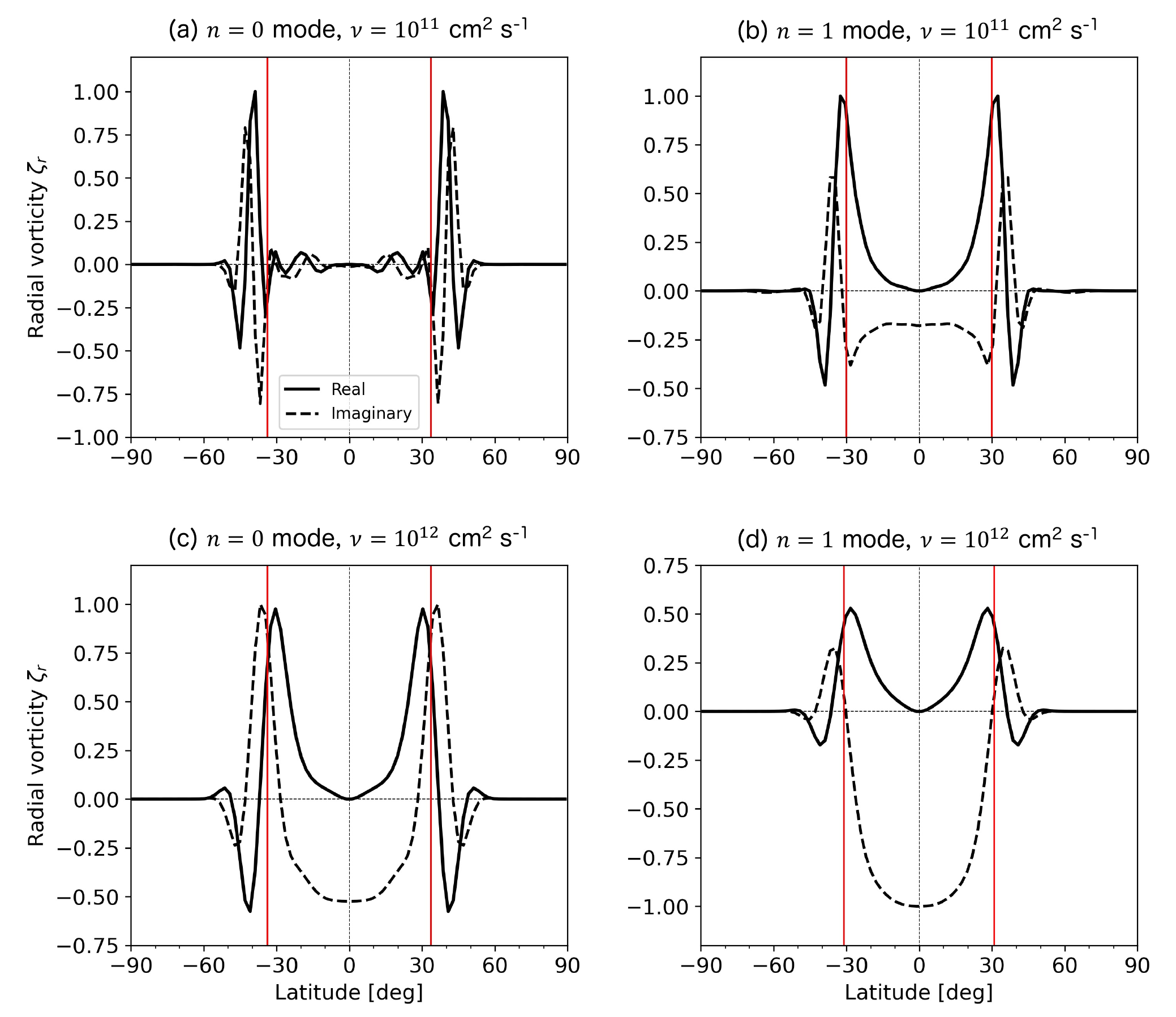}
     \caption{
     Radial vorticity $\zeta_{r}$ eigenfunctions at the surface ($r=0.985R_{\odot}$) for $m=8$.
     Left and right panels show the cases for the equatorial Rossby modes with no radial nodes ($n=0$) and with one radial node ($n=1$).
     Upper and lower panels show the cases with weak diffusion ($\nu=10^{11}$ cm$^{2}$ s$^{-1}$) and strong diffusion ($\nu=10^{12}$ cm$^{2}$ s$^{-1}$).
     Black solid and dashed lines represent real and imaginary eigenfunctions, respectively.
     The real part of the eigenfunctions are defined to be zero at the equator.
     The vertical red lines denote the location of critical latitudes where the phase speed of a Rossby mode is equal to the differential rotation velocity.
     }
     \label{fig:wr1d_dr}
   \end{figure*}
%___________________________________________________________

%___________________________________________________________
   \begin{figure*}[]
     \centering
     \includegraphics[width=0.85\linewidth]{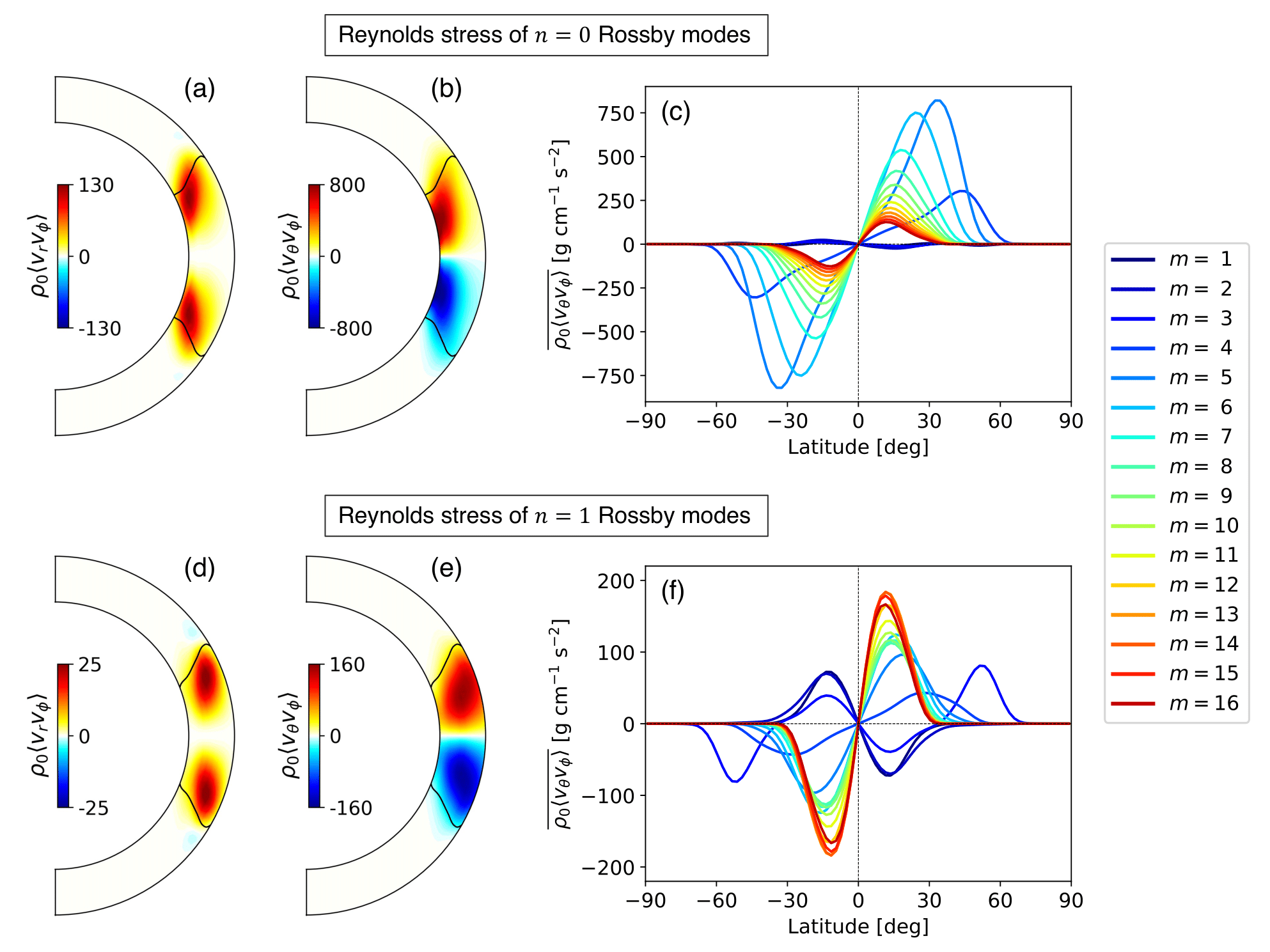}
     \caption{Reynolds stress components (a)(d) $\rho_{0}\langle v_{r}v_{\phi}\rangle$ and (b)(e) $\rho_{0}\langle v_{\theta}v_{\phi}\rangle$ associated with the equatorial Rossby modes for $m=8$.
     The units are g~cm$^{-1}$~s$^{-2}$.
     Black solid lines denote the location of critical latitudes at each height.
     The eigenfunctions are normalized such that the maximum horizontal velocity at the top boundary is $2$ m~s$^{-1}$.
     Panels (c) and (f) show the horizontal Reynolds stress averaged over radius $\overline{\rho_{0}\langle v_{\theta}v_{\phi}\rangle}$ where the overbar denotes the radial average.
     Different colors represent different azimuthal orders.
     Upper and lower panels show the cases for $n=0$ modes and $n=1$ Rossby modes, respectively.
     }
     \label{fig:RS_eqRos_dr}
   \end{figure*}
%___________________________________________________________

%___________________________________________________________
   \begin{figure}%[h!]
     \centering
     \includegraphics[width=\linewidth]{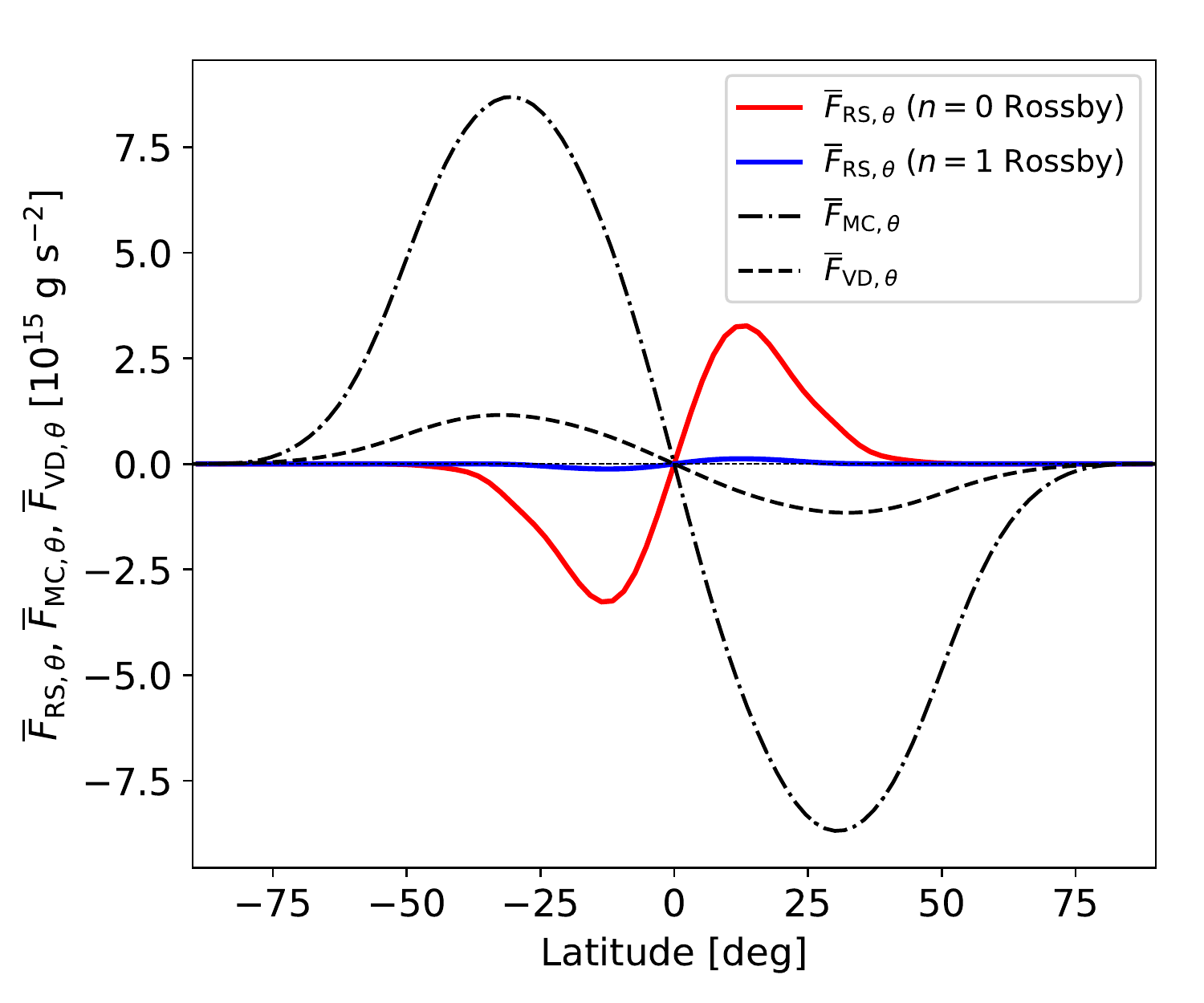}
     \caption{
     Radially averaged latitudinal angular momentum fluxes.
     Solid, dot-dashed, and dashed lines represent the angular momentum fluxes associated with the Reynolds stress of the equatorial Rossby modes $F_{\mathrm{RS,\theta}}$, advection by meridional circulation $F_{\mathrm{MC,\theta}}$, and diffusion by turbulent viscosity $F_{\mathrm{VD,\theta}}$, defined by the Eqs.~(\ref{eq:F_RS})--(\ref{eq:F_VD}).
     Red and blue lines represent the equatorial Rossby modes with no radial nodes ($n=0$) and with one radial node ($n=1$), respectively.
     Here, eigenfunctions are normalized such that the maximum horizontal velocity amplitude at the surface is $2$ m~s$^{-1}$.
     }
     \label{fig:gyro_dr}
   \end{figure}
%___________________________________________________________

%__________________________________________________________________
%__________________________________________________________________

\section{Effect of non-adiabatic stratification} \label{sec:delta}

In this section, the effects of non-adiabatic stratification are investigated.
While theoretical model of the Sun conventionally assume a slightly positive superadiabaticity value $0<\delta \lesssim 10^{-6}$ \citep[e.g.,][]{ossendrijver2003},
recent numerical simulations of solar convection imply that the lower half of the convection zone might be slightly subadiabatic \citep[][]{hotta2017,kapyla2017,bekki2017b,karak2018,kapyla2019}.
To this end, we vary the superadiabaticity from weakly subadiabatic to weakly superadiabatic, $\delta=-2\times 10^{-6},-10^{-6},0,10^{-6},2\times 10^{-6}$, while keeping the diffusivities fixed ($\nu=\kappa=10^{12}$ cm$^{2}$ s$^{-1}$).
The solar differential rotation and latitudinal entropy gradient are not included.
Since the entropy perturbation is generated by the radial flow, in this section, we focus on the (north-south $\zeta_z$-symmetric) columnar convective modes where strong radial motions are involved.

Figure~\ref{fig:eigfreq_thRos}a shows the dispersion relations of the $\zeta_z$-symmetric columnar convective modes for different $\delta$.
As the background becomes more subadiabatic (superadiabatic), the mode frequencies become higher (lower), i.e., the modes propagate in a prograde direction with faster (slower) phase speed.
When $\delta$ is sufficiently large, the imaginary mode frequencies become positive, i.e., the modes become convectively unstable.
This is clearly manifested in Fig.~\ref{fig:eigfreq_thRos}b where the mode frequencies are plotted in a complex plane.
Each points denote each mode with the associated azimuthal order labelled nearby.
The stable and unstable modes are distinguished by circles and diamonds, respectively.
For $\delta>0$ (blue and purple), a sudden transitions occurs from stable to unstable branches (at $m=5$ and $4$).
The critical azimuthal order for this transition depends on the superadiabaticity $\delta$ via the Rayleigh number criterion for the convective instability.

The changes in the dispersion relation can be understood by considering whether the buoyancy force acts as a restoring force or the opposite.
Figures~\ref{fig:delta_thRos}a and~b present the snapshots of $v_{r}$ and $s_{1}$ in an equatorial plane seen from the north pole for weakly subadiabatic background ($\delta=-2\times 10^{-6}$) for $m=3$ and $8$, respectively.
It is seen that the phase with positive $s_{1}$ is always ahead of the phase with positive $v_{r}$ in longitude, leading to a negative correlation between $\Re[v_{r}]$ and $\Im[s_{1}]$.
This physically means that, in this case, the buoyancy force acts as an additional restoring force for prograde-propagating columnar convective modes.
In other words, these modes share a property of prograde-propagating g modes.
Consequently, the mode frequencies become higher for $\delta<0$.
The opposite situation happens for $\delta>0$.
Figures~\ref{fig:delta_thRos}c and~d show the same equatorial cuts of $v_{r}$ and $s_{1}$ for weakly superadiabatic background.
When $m$ is not large enough for the convective instability to occur, it is seen that the phase with positive $s_{1}$ is behind the phase with positive $v_{r}$ in longitude, leading to a positive correlation between $\Re[v_{r}]$ and $\Im[s_{1}]$.
Therefore, the buoyancy force acts against the original restoring force of the compressional $\beta$-effect, which weakens the prograde propagation of columnar convective modes.
As a consequence, the mode frequencies become lower for $\delta>0$.
This effect was first studied in \citet[][]{gilman1987} using a simplified cylindrical model.
Figure~\ref{fig:delta_thRos}d shows the case where $m$ is sufficiently large and the mode becomes convectively unstable.
It is obviously seen that the phases of $v_{r}$ and $s_{1}$ now coincide and they both have the same sign at each phase, leading to $\langle v_{r}s_{1}\rangle>0$.

Figure~\ref{fig:FeRS_thRos} further shows the transport properties of thermal energy and angular momentum by convectively-unstable columnar convective modes.
Shown is the case with $\delta=2\times 10^{-6}$ and for $m=16$.
Positive $\langle v_{r}T_{1}\rangle$ in Fig.~\ref{fig:FeRS_thRos}a manifests that the enthalpy flux is transported upward.
The Reynolds stress components $\langle v_{r}v_{\phi}\rangle$ and $\langle v_{\theta}v_{\phi}\rangle$ are representatives of the radial and latitudinal angular momentum fluxes, respectively.
It is shown that the columnar convective modes can transport the angular momentum radially upward in the bulk of the convection zone and eqautorward near the surface.
This agrees with the results found in the rotating convection simulation (Bekki et al., in prep.).

%__________________________________________________________________
%__________________________________________________________________
%__________________________________________________________________
%__________________________________________________________________

\section{Effect of solar differential rotation} \label{sec:diffrot}

Finally, in this section, we take into account the effects of solar differential rotation.
For prescribing $\Omega(r,\theta)$, we use the data obtained from global helioseismology inversions from MDI and HMI \citep{larson2018} as shown in Fig.~\ref{fig:dr}a.
Note that the observational data is truncated at $r=r_{\mathrm{min}}$ and $r_{\mathrm{max}}$, and therefore, the effects of strong radial shear layers such as tachocline and the near surface shear layer of the Sun are not included.
The observing frame is chosen to be Carrington frame with the rotation rate $\Omega_{0}/2\pi=456.0$ nHz.
Figure~\ref{fig:dr}b shows the latitudinal profiles of the differential rotation at different depths.
Horizontal dashed lines indicate the estimated phase speed of the $n=0$ equatorial Rossby modes, $-2\Omega_{0}/\left[m(m+1)\right]$, for selected $m$ values.
For $m>2$, there emerge critical latitudes where the phase speed of the Rossby mode matches with the differential rotation speed.
As discussed in \citet{gizon2020} and \citet{fournier2022}, turbulent viscous diffusion is required to get rid of the singularities at the critical latitudes, leading to a formation of viscous critical layers with the typical latitudinal extent $\delta_{\mathrm{crit}}$ given by
\begin{eqnarray}
&& \frac{\delta_{\mathrm{crit}}}{R_{\odot}}\approx \left( \frac{\nu}{m\Omega_{0}R_{\odot}^{2}}\right)^{1/3}.
\end{eqnarray}

Figure~\ref{fig:freq_dr} shows the distribution of eigenfrequencies of the global-scale inertial modes in a complex plane for $m=2$ and $16$.
Shown in shaded area represent the range of mode frequencies where differential rotation can have a strong impact by producing the critical layers.
As higher $m$, the number of eigenmodes that are affected by differential rotation increases:
In fact, most of the retrograde-propagating inertial modes are affected by critical latitudes at higher $m$ (see Fig.~\ref{fig:freq_dr}b).

%__________________________________________________________________
%__________________________________________________________________

\subsection{Rossby modes with viscous critical layers} \label{sec:viscouscrit}

In this section, we carry out a set of calculations for $\nu=10^{11}$ and $10^{12}$ cm$^{2}$ s$^{-1}$ with the differential rotation included to study how the equatorial Rossby modes are affected by the viscous critical layers.
For the sake of simplicity, the background is set to be perfectly adiabatic and the latitudinal entropy variation $\partial s_{0}/\partial \theta$ is switched off.

Figure~\ref{fig:disp_eqRos_dr}a shows the dispersion relation of the equatorial Rossby modes with $n=0$ (red) and $n=1$ (blue) for weak (dashed) and strong (solid) viscous diffusivities, respectively.
Shown in white circles, squares, and diamonds are the frequencies of the Rossby modes observed on the Sun \citep{loeptien2018,liang2019,proxauf2020}.
The viscous diffusivity value is found to have a rather small effect on the real part of their eigenfrequencies. 
At $m=3$, the observed frequency agrees almost perfectly with the $n=0$ equatorial Rossby mode's frequency.
However, for $m>3$, the observed frequencies lie in between the frequencies of $n=0$ and $n=1$ modes.
Figure~\ref{fig:disp_eqRos_dr}b shows the computed eigenfrequencies in a complex plane.
Unlike the $n=1$ modes, the $n=0$ modes are substantially damped only for $m \geq 4$, which is likely owing to the emergence of the critical latitudes that significantly modify the $n=0$ modes' eigenfunctions.
The linewidths of the equatorial Rossby modes in our model are of the same order of magnitudes as the observations as shown in Fig.~\ref{fig:disp_eqRos_dr}b.

Figure~\ref{fig:eigenfunc_eqRos_dr} shows the velocity eigenfunctions of the $n=0$ modes for the case with $\nu=10^{12}$ cm$^{2}$ s$^{-1}$.
Figures~\ref{fig:eigenfunc_eqRos_dr}a and c show meridional cuts through the eigenfunctions for $m=5$ and $12$, respectively.
As already discussed in \S~\ref{sec:viscosity}, the latitudinal velocity is confined close to the base of the convection zone.
With differential rotation included, they are further trapped in the equatorial region bounded by the viscous critical layers. 
Unlike the uniformly-rotating case, strong radial and longitudinal flows are driven around the critical latitudes, which leads to strong concentrations of $z$-vorticity there. 
Figures~\ref{fig:eigenfunc_eqRos_dr}b and~d show the latitudinal velocity $v_{\theta}$ (top rows) and radial vorticity $\zeta_{r}$ (bottom rows) at the top of the domain.
They both have a similar chevron-like inclination towards the equator.
However, $\zeta_{r}$ has prominent power peaks around the critical layers.

Figure~\ref{fig:eigenfunc_n1eqRos_dr} is the same figure as Fig.~\ref{fig:eigenfunc_eqRos_dr} but for the $n=1$ equatorial Rossby modes.
Unlike the $n=0$ modes, the eigenfunctions of $v_{\theta}$ (and $\zeta_{r}$) peak at the surface and at the equator.
Although the critical layers exist similarly to the $n=0$ modes, they are found to have a rather limited impact on the $n=1$ Rossby modes.

To see the diffusivity dependence, we show $\zeta_{r}$ at the surface  for weak (top rows) and strong viscous diffusivities (bottom rows) in Fig.~\ref{fig:wr1d_dr}.
The left and right panels are for the $n=0$ and $n=1$ equatorial Rossby modes, respectively.
The solid and dashed lines denote the real and imaginary parts, and the vertical red line indicates the location of the critical latitudes.
The phase is defined such that $\Re[\zeta_{r}]=0$ at the equator and the maximum amplitudes are normalized to unity.
Substantial structure is observed associated with the viscous critical layers.
In general, this structure becomes broader and weaker as the viscosity $\nu$ is increased.
It is also seen that amplitudes of the imaginary parts of the eigenfunctions are larger for $n=0$ modes than for $n=1$ modes.

Next, let us examine the impact of the net angular momentum transport by the equatorial Rossby modes under the influences of solar differential rotation.
Figures~\ref{fig:RS_eqRos_dr}a and~b show the Reynolds stress components $\langle v_{r}v_{\theta} \rangle$ and $\langle v_{\theta}v_{\theta} \rangle$ for $n=0$ modes at $m=8$.
The Reynolds stresses become substantially non-zero near the viscous critical layers.
It is striking that even $n=0$ mode, which in the case of uniform rotation is toroidal and non-convective, can transport the angular momentum radially upward around the viscous critical layers. 
Latitudinally, the angular momentum is transported equatorward in both hemispheres.
Figure~\ref{fig:RS_eqRos_dr}c shows the $\langle v_{\theta}v_{\theta} \rangle$ at the surface for all $m$.
It is seen that the correlations become small as $m$ increases because the $n=0$ modes are more and more confined closer to the base of the convection zone.
The counterparts for $n=1$ modes are shown in  Fig.~\ref{fig:RS_eqRos_dr}d--f. 
It is clear that the $n=1$ modes also transport angular momentum radially upward and equatorward at higher $m$.
However, unlike the $n=0$ modes, the Reynolds stress $\langle v_{\theta}v_{\theta} \rangle$ peaks slightly below the surface.
Therefore, the correlation at the surface becomes more prominant as $m$ increases, as shown in Fig.~\ref{fig:RS_eqRos_dr}f.

It is instructive to examine how significant the angular momentum transport by these equatorial Rossby modes can be in the Sun.
To this end, we consider the so-called gyroscopic pumping equation \citep[e.g.,][]{elliott2000,miesch2011}
\begin{eqnarray}
&& \nabla\cdot\left( \bm{F}_{\mathrm{RS}}+\bm{F}_{\mathrm{MC}}+\bm{F}_{\mathrm{VD}} \right)=0, \label{eq:gyro}
\end{eqnarray}
where $\bm{F}_{\mathrm{RS}}$, $\bm{F}_{\mathrm{MC}}$, and $\bm{F}_{\mathrm{VD}}$ are the angular momentum fluxes transported by the Reynolds stress, meridional circulation, and turbulent viscous diffusion, respectively.
They are defined by
\begin{eqnarray}
&& \bm{F}_{\mathrm{RS}}=\rho_{0}r\sin{\theta}\ \langle   v_{\phi} \bm{v}_{\rm m}\rangle, \label{eq:F_RS} \\
&& \bm{F}_{\mathrm{MC}}=\rho_{0} (r\sin{\theta})^2    \Omega\ \bm{v}_{\rm m}, \label{eq:F_MC} \\
&& \bm{F}_{\mathrm{VD}}=-\rho_{0}\nu (r\sin{\theta})^2\     \nabla\Omega, \label{eq:F_VD} 
\end{eqnarray}
where $\bm{v}_{\rm m}$ is the meridional flow.
Figure~\ref{fig:gyro_dr} shows the each term of the latitudinal component of the Eq.~(\ref{eq:gyro}) averaged over radius.
The eigenfunctions are normalized such that the maximum horizontal velocity amplitude at the surface is $2$ m~s$^{-1}$, as inferred from observations \citep{loeptien2018}.
To estimate $F_{\mathrm{MC},\theta}$ (black dot-dashed line), we use the observational meridional circulation data obtained by \citet{gizon2020s}.
For $F_{\mathrm{VD},\theta}$ (black dashed line), we assume the spatially-uniform viscosity of $\nu=10^{12}$ cm$^{2}$ s$^{-1}$.
It is shown that the equatorward angular momentum transport by the Reynolds stress is balanced by the poleward transport by meridional flow and by turbulent diffusion.
The amplitude of $F_{\mathrm{RS},\theta}$ associated with $n=1$ modes are found to be almost negligible, whereas that of $n=0$ modes is substantial and accounts for about $30-40$ \% of the other two contributions $F_{\mathrm{MC},\theta}+F_{\mathrm{VD},\theta}$.
The difference between the $n=0$ and $n=1$ modes comes from that fact that the velocity eigenfunctions of the $n=1$ modes peak at the surface, whereas those of the $n=0$ modes peak near the base.
Therefore, when the eigenfunctions are normalized by the surface velocity speed, only $n=0$ modes become important for the convection zone dynamics.
Some caution must be given here as the eigenfunctions can also be highly sensitive to various model parameters (such as $\nu$ and $\delta$), and thus, a different set of parameters might lead to a different angular momentum balance.
Furthermore, the model assumes that the diffusivities are uniform and isotropic, which will also affect the eigenfunctions.
Nonetheless, it is suggested that the equatorial Rossby modes might potentially play a role in transporting the angular momentum equatorward in the Sun.

%___________________________________________________________
   \begin{figure*}[]
     \centering
     \includegraphics[width=\linewidth]{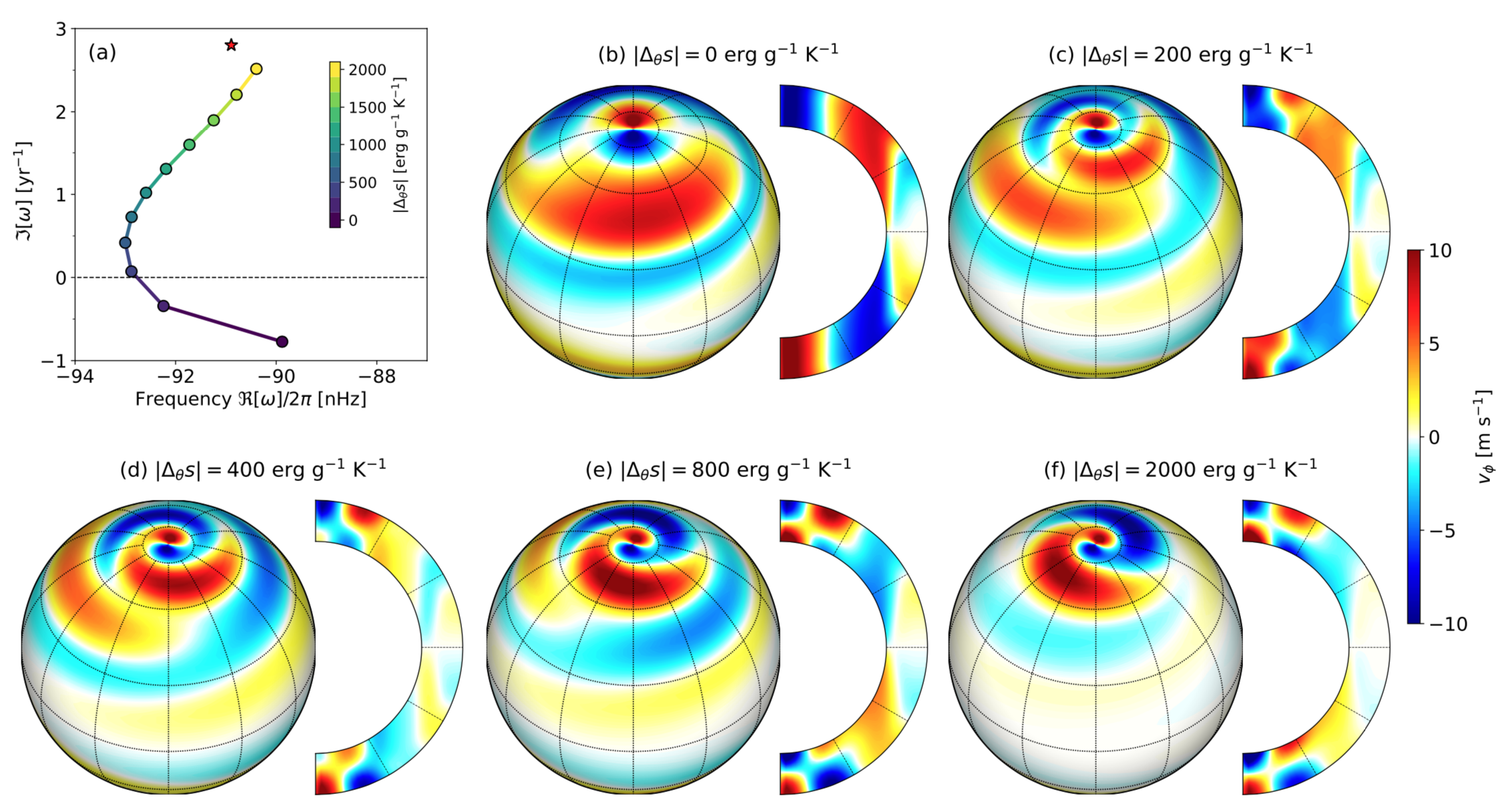}
     \caption{
     (a) Growth rate versus frequency of the  $\zeta_z$-antisymmetric high-latitude mode with $m=1$, for different values  of the latitudinal entropy difference $|\Delta_{\theta} s| = s_{0,{\rm pole}}-s_{0,{\rm eq}}$. 
     The red star is for the case of a realistic latitudinal entropy gradient that depends on position, estimated according to Eq.~(\ref{eq:twb}).
     Realistic  solar differential rotation is included.
     The background stratification is adiabatic ($\delta=0$) and the diffusivities are set to $\nu=\kappa=10^{12}$ cm$^{2}$ s$^{-1}$.
     (b)-(f) Eigenfunctions of the longitudinal velocity $v_{\phi}$ at the surface ($r=0.985R_{\odot}$) and at the central meridian for some selected $\Delta s_{\theta}$.
     The eigenfunctions are normalized so that the maximum flow amplitudes at the surface is $v_{\phi}=10$ m~s$^{-1}$. 
     }
     \label{fig:tpgRos_m1}
   \end{figure*}
%___________________________________________________________

%___________________________________________________________
   \begin{figure*}[]
     \centering
     \includegraphics[width=0.765\linewidth]{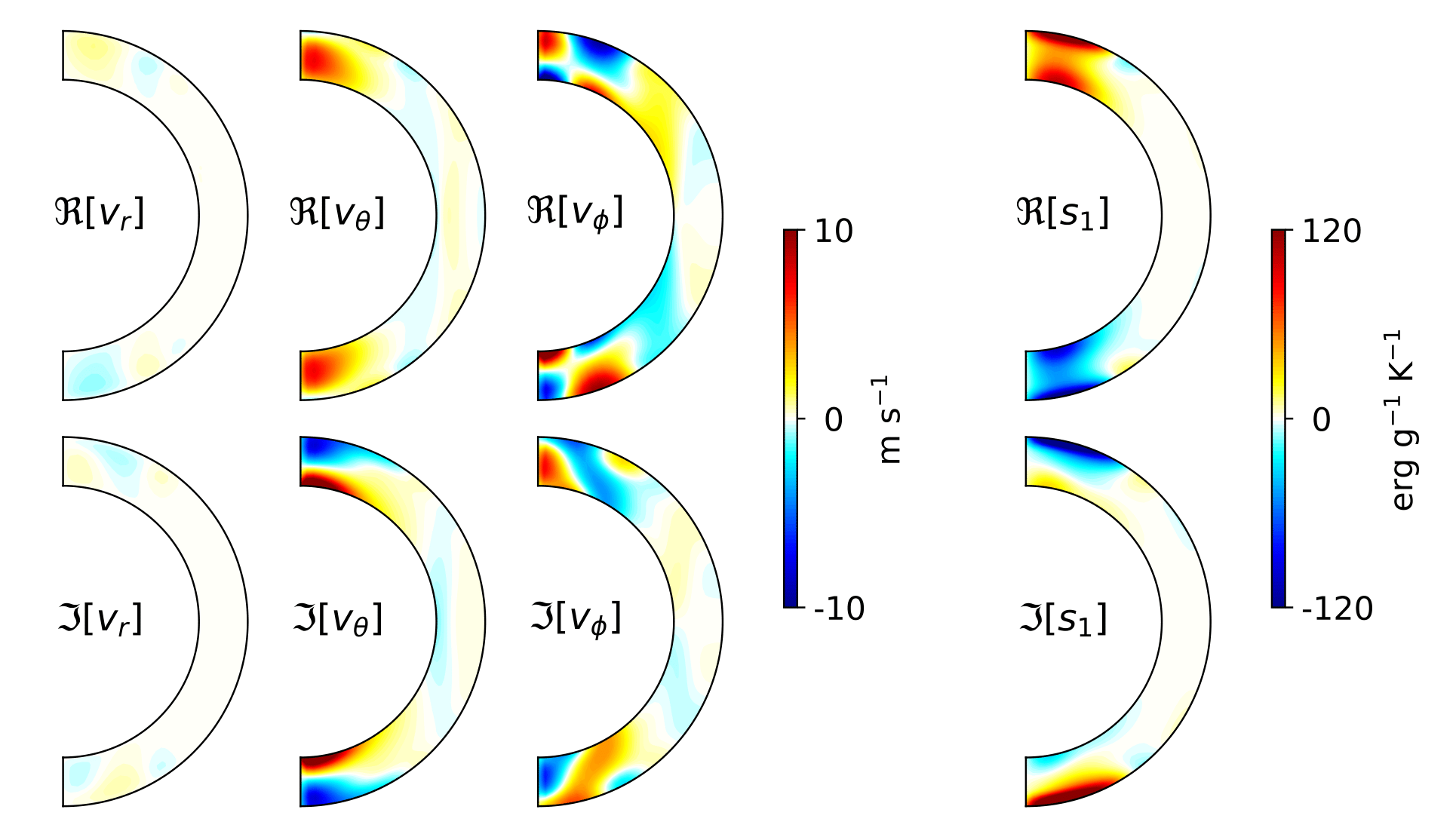}
     \caption{
     Eigenfunctions of $v_{r}$, $v_{\theta}$, $v_{\phi}$, and $s_{1}$ of the $m=1$ north-south $\zeta_z$-antisymmetric high-latitude inertial mode with the solar differential rotation and the corresponding latitudinal entropy gradient (Eq.~\ref{eq:twb}) included.
     The background stratification is adiabatic ($\delta=0$) and the diffusivities are set to $\nu=\kappa=10^{12}$ cm$^{2}$ s$^{-1}$.
     The eigenfunctions are normalized such that the maximum $v_{\phi}$ is $10$ m~s~$^{-1}$ at the surface.
     }
     \label{fig:eig_highlat_m1}
   \end{figure*}
%___________________________________________________________

%__________________________________________________________________
%__________________________________________________________________

\subsection{Effect of baroclinicity on high-latitude inertial modes}\label{sec:highlat_m1}

We assume the solar differential rotation is in thermal wind balance;
where the deviation from the Taylor-Proudman's state is balanced by the latitudinal entropy variation \citep[e.g.,][]{rempel2005,miesch2006,brun2011}.
In other words, the solar convection zone is essentially baroclinic.
Since the high-latitude modes are located at high latitudes, they are subject to the imposed baroclinicity in the convection zone and potentially become unstable \citep[][]{knobloch1982,spruit1984,kitchatinov2013,gilman2014}.

In this section, we study the effect of baroclinicity in the convection zone on the high-latitude inertial modes by varying the amplitude of the imposed latitudinal entropy gradient.
Of particular interest is this effect on the $m=1$ mode with north-south antisymmetric $\zeta_{z}$. 
Here, we assume the latitudinal dependence of the background entropy profile is
\begin{eqnarray}
&&  \frac{\partial s_{0}}{\partial\theta}= -|\Delta_\theta s| \ \sin{2\theta},
\end{eqnarray}
where $\Delta_{\theta}s =s_{0, \rm eq}-s_{0, \rm pole} (< 0)$ represents the entropy difference between the cooler equator and the hotter poles.
For simplicity, the radial dependence is ignored ($s_{0}$ is uniform in radius and thus convectively neutral).
We use moderately viscous and thermal diffusivities $\nu=\kappa=10^{12}$ cm$^{2}$ s$^{-1}$.

Figure~\ref{fig:tpgRos_m1}a shows the eigenfrequencies of the $m=1$ north-south $\zeta_z$-antisymmetric high-latitude modes in a complex plane with varying $|\Delta_{\theta} s|$ from $0$ to $2000$ erg g$^{-1}$ K$^{-1}$.
It is shown that, as the baroclinicity is increased, the modes become unstable ($\Im[\omega]>0$).
In this sense, these modes can also be called baroclinic (Rossby) modes.
Figures~\ref{fig:tpgRos_m1}b--f show the eigenfunctions of $v_{\phi}$ both at the surface and at the central meridian for different values of $\Delta_{\theta}  s$.
It is clearly seen that, as $|\Delta_{\theta} s|$ increases and the high-latitude mode becomes more and more baroclinically unstable, it begins to exhibit a spiralling flow structure around the poles.
The spatial extent and the tilt of this spiral agree strikingly well with the observations, see \citet{hathaway2020} and \citet{gizon2021}.

In order to assess if the baroclinicity in the Sun is large enough for the baroclinic instability to occur, we estimate the latitudinal entropy variation using the helioseismically-constrained differential rotation profile using,
\begin{eqnarray}
&& \frac{g}{\cp}\frac{\partial s_{0}}{\partial\theta}=r^{2}\sin{\theta} \ \frac{\dd (\Omega^{2}) }{\dd z} . \label{eq:twb}
\end{eqnarray}
With this realistic baroclinicity included in our model (Eq.~\ref{eq2:ent}), 
we find that the $m=1$ high-latitude mode is self-excited:
The growth rate is $\Im[\omega]/2\pi=14.1$ nHz, which translates into the growing time scale of $4.3$ months.
This may explain why the high-latitude flow feature on the Sun has a much larger flow amplitude than the equatorial Rossby modes.
Its mode frequency is $\Re[\omega]/2\pi=-90.9$ nHz (measured in the Carrington frame), which is close to the observed propagation frequency of the high-latitude flow feature of $-86.3$ nHz \citep{gizon2021}.
The eigenfunctions of this $m=1$ mode are shown in Fig.~\ref{fig:eig_highlat_m1}.
The mode is characterized by its dominant $z$-vortical motion and is quasi-toroidal (the vertical flow is about $10$ times weaker than the horizontal ones).
It is clearly seen that, unlike the case without baroclinicity, a strong entropy perturbation is associated with this mode.
The mode is strongly localized near the poles.
Although this $m=1$ mode has successfully been detected on the Sun, we find that about $30\%$ of the total kinetic energy exists above $80^{\circ}$ latitude, which is the observational upper limit of the ring-diagram measurements presented by \citet[][]{gizon2021}.
This means that the observations may miss a  fraction of the mode power.

Using a linear model of Boussinesq convection in a uniformly-rotating 
spherical shell, \citet[][]{gilman1975} found convectively-unstable modes near the poles with a spiral structure (see his figure~17). 
Like the high-latitude modes we have found, his modes lie mostly inside the tangent cylinder. 
However, his modes are convectively driven, whereas ours are baroclinically driven due to the solar differential rotation and the latitudinal entropy variation.

%__________________________________________________________________
%__________________________________________________________________

%__________________________________________________________________
%__________________________________________________________________

%\newpage

\section{Summary} \label{sec:summary}

In this paper, we have presented a linear modal analysis of the oscillations of the solar convection zone at low frequencies.
We have reported dispersion relations and eigenfunctions of the equatorial Rossby modes without a radial node ($n=0$) and with one radial node ($n=1$), and the columnar convective modes and the high-latitude modes, both with different north-south symmetries.
We find ``mixed Rossby modes'' which share properties of the $n=1$ equatorial Rossby modes and the $\zeta_z$-antisymmetric columnar convective modes.

%Our main findings are summarized as follows.

We studied the effects of the turbulent diffusion and the solar differential rotation on the equatorial Rossby modes.
Our main findings are summarized as follows.
One effect of turbulent diffusion is to radically change the radial force balance of the $n=0$ equatorial Rossby modes. 
The modes are confined closer to the base of the convection zone and their eigenfunctions deviate strongly from the well-known $r^{m}$ radial dependence.
When the solar differential rotation is taken into account, viscous critical layers are formed in latitudes where the phase speed of the equatorial Rossby mode is equal to the differential rotation speed. 
Strong radial and longitudinal flows are present in the viscous critical layers and the eigenfunctions are complex, implying non-zero Reynolds stresses.
We also find that, unlike the $n=0$ equatorial Rossby modes, the ``mixed modes'' are almost unaffected by the presence of solar differential rotation and strong viscous diffusivity.
The retrograde frequencies of the observed Rossby modes of the Sun have values in between the model eigenfrequencies of the $n=0$ and $n=1$ modes for $m \geq 5$ (see Fig.~\ref{fig:disp_eqRos_dr}a).

We have further demonstrated that the dispersion relations of the columnar convective modes are very sensitive to the background superadiabaticity $\delta$.
These modes are convectively unstable and transport thermal energy and angular momentum when $\delta>0$.
We have also shown that the $m=1$ high-latitude mode is substantially modified by a latitudinal entropy gradient.
When the latitudinal entropy gradient exists, the mode is baroclinically unstable and the eigenfunction at the surface matches the observations with the correct geometry, including the correct sense for the spiral seen in $v_\phi$.
These results imply that the observations of the solar inertial modes can be used to measure the degree of non-adiabaticity in the Sun, as proposed by \citet[][]{gilman1987}.

Several simplifying assumptions were made in this study.
For instance, the viscous and thermal diffusivities, $\nu$ and $\kappa$, and the superadiabaticity $\delta$ were all assumed to be spatially uniform, which is not realistic. 
Moreover, we set the bottom and top boundaries at $(r_{\mathrm{min}},r_{\mathrm{max}})=(0.71R_{\odot},0.985R_{\odot})$ and thus both the tachocline and the near-surface shear layer of the Sun were excluded from our model.
Future work will be to include the radiative interior and the photosphere and to allow for a radial dependence of $\nu$, $\kappa$ and $\delta$. 
In addition, it will be important to compare the present results to modes extracted from three-dimensional numerical simulations of rotating convection in the strongly non-linear regime.
The aim is to have a physical understanding of all the modes in the low-frequency spectrum and thus a reliable identification of the observed modes, including the critical-latitude modes.

%%%%%%%%%%%%%%%%%%%%%%%%%%%%%%%%%%%%%%%
%%%%%%%%%%%%%%%%%%%%%%%%%%%%%%%%%%%%%%%
\begin{acknowledgements}
%\YB{The authors thank an anonymous referee for insightful comments on the manuscript.}
We thank A.C.~Birch and B.~Proxauf for helpful discussions. 
YB did most of the work, RHC and LG provided supervision.
YB is a member of the International Max-Planck Research School for Solar System Science at the University of G\"ottingen. 
We acknowledge support from ERC Synergy Grant WHOLE SUN 810218 and the hospitality of the Institut Pascal in March 2022.
YB is the beneficiary of a long-term scholarship program for degree-seeking graduate students abroad from the Japan Student Services Organization (JASSO).
\end{acknowledgements}

%-------------------------------------------------------------------
%----------------------- reference list-----------------------------
\bibliographystyle{aa} % style aa.bst
\bibliography{ref} % your references Yourfile.bib
%-------------------------------------------------------------------

%%%%%%%%%%%%%%%%%%%%%%%%%%%%%%%%%%%%%%%
%%%%%%%%%%%%%%%%%%%%%%%%%%%%%%%%%%%%%%%

%\Online
\clearpage
\newpage
\begin{appendix} %First online appendix

\end{appendix}

\end{document}